\pdfoutput=1
\documentclass[12pt,reqno]{amsart}
\usepackage{amsmath,amsfonts,amssymb,amscd,amsthm,amsbsy, color}
\usepackage[pdftex]{graphicx}
\usepackage{tensor}
\usepackage{comment}
\usepackage[format=plain,
justification=raggedright,singlelinecheck=false]{caption}
\usepackage[maxfloats=100]{morefloats}
\usepackage{courier}
\usepackage{wrapfig}
\usepackage{fancybox}

\usepackage{multimedia}
\usepackage{graphicx,epsfig}

\usepackage{appendix}
\usepackage{natbib}

\defcitealias{mL89}{Lane's}
\defcitealias{gG74}{Giacaglia's}

\numberwithin{equation}{section}

\newtheorem{theorem}{Theorem}
\newtheorem{meta-thm}[theorem]{Meta-Theorem}

\newtheorem{definition}[theorem]{Definition}

\newcommand\beq[1]{ \begin{equation}\label{#1} }
\newcommand{\eeq}{ \end{equation} }

\newcommand\beqa[1]{ \begin{eqnarray} \label{#1}}

\newcommand{\eeqa}{ \end{eqnarray} }
\newcommand{\beqano}{ \begin{eqnarray*} }
\newcommand{\eeqano}{ \end{eqnarray*} }
\newcommand\equ[1]{{\rm (\ref{#1})}}

\def\G{{\mathcal G}}
\def\H{{\mathcal H}}

\def\integer{{\mathbb Z}}

\def\real{{\mathbb R}}

\begin{document}

\title[A study of the lunisolar secular resonance $2\dot{\omega}+\dot{\Omega}=0$]
{A study of the lunisolar secular resonance $2\dot{\omega}+\dot{\Omega}=0$}

\author[A. Celletti]{Alessandra Celletti}
\address{
Department of Mathematics, University of Roma Tor Vergata, Via della Ricerca Scientifica 1,
00133 Roma (Italy)}
\email{celletti@mat.uniroma2.it}

\author[C. Gale\c s]{C\u at\u alin Gale\c s}
\address{
Department of Mathematics, Al. I. Cuza University, Bd. Carol I 11,
700506 Iasi (Romania)}
\email{cgales@uaic.ro}

\thanks{A.C. was partially supported by the European Grant MC-ITN Stardust,
PRIN-MIUR 2010JJ4KPA$\_$009 and GNFM/INdAM;
C.G. was supported by a grant of the Romanian National Authority for
Scientific Research and Innovation, CNCS - UEFISCDI, project number
PN-II-RU-TE-2014-4-0320 and by GNFM/INdAM.}

\date{Received: date / Accepted: date}

\begin{abstract}
The dynamics of small bodies around the Earth has gained a renewed interest,
since the awareness of the problems that space debris can cause in the
nearby future. A relevant role in space debris is played by lunisolar
secular resonances, which might contribute to an increase of the orbital
elements, typically of the eccentricity. We concentrate our attention on
the lunisolar secular resonance described by the relation
$2\dot{\omega}+\dot{\Omega}=0$, where $\omega$ and $\Omega$ denote
the argument of perigee and the longitude of the ascending node of the space debris.
We introduce three different models with increasing complexity. We show that the growth in eccentricity, as observed in space debris located in the MEO region at the inclination about equal to $56^\circ$, can be explained as a natural effect of the secular resonance $2\dot{\omega}+\dot{\Omega}=0$, while the chaotic variations of the orbital parameters are the result of interaction and overlapping of nearby resonances.
\end{abstract}

\keywords{Space debris, Lunisolar secular resonance, Eccentricity growth}


\maketitle

\section{Introduction}
Thousands of man-made objects, abandoned during space missions or remnants of operative satellites, orbit
around the Earth at different altitudes. Their size varies from larger pieces, like old satellites
or rocket stages, to dust-size particles given by fragmentation of satellites or even by collision
events, like the impact between Kosmos 2251 and Iridium 33 in 2009, or the destruction of Fengyun-1C in 2007.

The dynamics of space debris strongly differs according to the altitude from the Earth. To this end,
one distinguishes 4 main regions as follows:

\begin{itemize}
    \item [$(i)$] the LEO (Low Earth Orbit) region spans the altitude from 0 to 2\,000 km;
    here the objects feel, in order of importance, the gravitational attraction of our planet,
    the dissipation due to the atmospheric drag, the Earth's oblateness effect,
    the attraction of Moon and Sun, and the solar radiation pressure;
    \item [$(ii)$] the MEO (Medium Earth Orbit) region goes from 2\,000 to 30\,000 km of altitude; the forces
    felt by the debris are like in LEO, except that there is no atmospheric drag;
    \item [$(iii)$] the GEO (Geostationary orbit) region is located around the value of 42\,164.17 km
    from the Earth's center; geostationary objects move with an orbital period equal to the rotational
    period of the Earth;
    \item [$(iv)$] HEO (High Earth orbit) region, refers to the space region with altitude above
    the geosynchronous orbit.
   \end{itemize}

In this work we are interested in a particular type of motion, which corresponds to a so-called
secular resonance. In particular, we consider the orbital elements which are solutions
of the relation
\beq{res1}
2\dot{\omega}+\dot{\Omega}=0\ ,
\eeq
where $\omega$ denotes the argument of perigee of the debris and $\Omega$ its longitude of the
ascending node. A relation like \equ{res1}, involving quantities moving on long time-scales,
is called a \sl secular resonance. \rm By considering the variations of $\omega$ and $\Omega$ as
just due to the effect of the main spherical harmonics of the geopotential, one can show that
equation \equ{res1} can be written just in terms of the inclination. As shown in \cite{HughesI},
there can be several secular resonances which depend on the inclination only. Among such resonances,
\equ{res1} represents a very interesting case, since it has been shown that it affects the dynamics
of objects in the MEO region (\cite{aR08}, \cite{Sanchez15}, \cite{RDGF2015}). Chaotic motions
arise from the interaction and overlapping of nearby resonances (\cite{aR15}, \cite{DRADVR15}, \cite{RDADRV15}).

In this paper we introduce three different models with increasing complexity, apt to study the
resonance \equ{res1}. The simplest model is described by a one degree-of-freedom autonomous Hamiltonian,
which is obtained by averaging over the fast angles and by neglecting the rates of variation of the
lunar longitude of the ascending node. This model provides the essential features, like the location
of stable equilibria with large as well as with small libration amplitude. The growth of the eccentricity can be easily explained by this integrable model. In the second model one does not average over the
fast angles, but still retains the assumption that the longitude of the ascending node of the Moon is
constant. Circulation and libration regions can be located, as well as the chaotic separatrix, although
the dynamics is very complicated: overlapping of resonances,  bifurcations and, as a consequence, the existence of equilibria
at large eccentricities as well as at small eccentricities, variation of the amplitude of the resonance. The last model includes the variation of the lunar longitude of the ascending node
and shows that large chaotic regions can appear, contributing to an irregular variation of the orbital elements.

\section{The model}\label{sec:model}
We consider a space debris subject to the gravitational attraction of the Earth, including the oblateness potential,
as well as the influence of Sun and Moon. This model is described by a Hamiltonian of the form
\begin{equation}\label{Hamiltonian}
\H=\H_{Kep}+\H_{Geo}+\H_{Moon}+\H_{Sun}\ ,
\end{equation}
which is the sum of different contributions that we are going to explain and express in terms of
the Delaunay action--angle variables $(L,G,H,M,\omega,\Omega)$, where the actions are defined by
\begin{equation}\label{Delaunay}
L=\sqrt{\mu_E a}\,, \quad G=L\sqrt{1-e^2}\,,\quad H=G \cos I\,,
\end{equation}
with $\mu_E= \G m_E$ the product of the gravitational constant $\G$ and the Earth's mass $m_E$, $a$ the semimajor axis, $e$ the orbital
eccentricity, $I$ the inclination, while the angle variables are the mean anomaly $M$, the argument of perigee $\omega$, the longitude of
the
ascending node $\Omega$, which are expressed with respect to the equatorial plane.

The first term in \equ {Hamiltonian} represents the Keplerian part $\H_{Kep}$, which can be expressed as
\beq{Hkep}
\H_{Kep}(L)=-\frac{\mu_E^2}{2 L^2}\ .
\eeq
The second term $\H_{Geo}$ describes the perturbation due to the Earth, when considering the shape of our planet.
In particular, we will consider only the most important term of the expansion in spherical harmonics of the geopotential,
the so-called $J_2$-term. Indeed, while studying the long--term dynamics of resonant orbits, the short--periodic terms
that depend on the mean anomaly of the satellite (as well as the mean anomaly of the perturbing body, when dealing with Sun and Moon)
can be averaged over from the disturbing function. Therefore, in the expression for  $\H_{Geo}$ we take an average
of the Hamiltonian over the mean anomaly of the space debris, which implies to consider only the most
important contribution, corresponding to the $J_2$ gravity coefficient of the secular part (see, e.g., \cite{CGmajor},
compare also with \cite{CGext}). This leads to express $\H_{Geo}$ in the form:
\beq{Hgeo}
\H_{Geo}(L,G,H)={{R_E^2 J_2 \mu_E^4}\over {4}}\ {{1}\over {L^3G^3}}\ (1-3{H^2\over G^2})\ ,
\eeq
where $R_E$ is the mean equatorial radius of the Earth and
$J_2=1.08263\times 10^{-3}$.

The contributions due to Moon and Sun are simplified by averaging over the fast angles, precisely the mean anomaly of the
debris and the mean anomalies of the perturbers (Moon and Sun). Moreover, we truncate the potentials to second order
in the ratio of semi-major axes (see \cite{wK62}, \cite{mL89} and \cite{CGPRnote} for details),
thus obtaining the expression for $\H_{Sun}$ and the (quite long) expression for $\H_{Moon}$,
reported in Appendix~\ref{app:rmoonsun} (see also \cite{gC62}).
Adding the contributions in \equ{Hkep}, \equ{Hgeo} as well as $\H_{Sun}$ and $\H_{Moon}$, we obtain the Hamiltonian \equ{Hamiltonian}.

Since the mean anomaly $M$ is a cyclic variable, its conjugated action $L$ (or equivalently the semi--major axis $a$) is constant.
As a consequence, the Hamiltonian system described by \eqref{Hamiltonian} is non--autonomous with two degrees of freedom.  As it was
remarked by \cite{aR15}, \cite{DRADVR15}, and analytically shown in \cite{CGPRnote}, the Hamiltonian $\H$ depends on time just through the
longitude of lunar ascending node $\Omega_{M}$ with a rate of variation equal to $\dot\Omega_M\simeq -0.053^{\circ}/day$,
which implies a periodicity of $\Omega_M$ over 18.6 years. More precisely, since $\dot\Omega_S =0$, where $\Omega_S$ is the longitude of the
solar ascending node, and the expansions of the lunar and solar potentials to second order in the ratio of semimajor axes are independent of
the lunar and solar perigees, it follows that $\H$ depends on time only through $\Omega_M$.

To a first approximation we assume that the Moon orbits on an
elliptic trajectory with semimajor axis equal to $a_M=384\,748$ km, eccentricity
$e_M=0.0549006$ and inclination $I_M=5^{\circ}15'$; the mass $m_M$
of the Moon, expressed in Earth's masses, is about equal to 0.0123. The orbital elements of the Moon are referred to the ecliptic plane.

As for the Sun, we can assume that its elements are constants and, precisely, $a_S=149\,597\,871$ km,
eccentricity $e_S=0.01671123$ and inclination $I_S=23^{\circ} 26' 21.406''$; the mass of the Sun $m_S$,
expressed in Earth's masses, is approximately equal to 333\,060.4016. The orbital elements of the Sun are expressed with respect to the
equatorial plane.

The model described by \equ{Hamiltonian} gives all the ingredients to capture the main dynamical features of the
resonant structure within the MEO region (see \cite{aR15} for a comparison between various models).

\section{The secular resonance $2\dot{\omega}+\dot{\Omega}=0$}\label{sec:res}

In this Section we are interested to the so-called (lunar and solar) \sl secular resonances, \rm which occur whenever
one has a commensurability between the arguments of perigee and the longitudes of the nodes
of the debris and the perturbers, according to the following definition.

\begin{definition} \label{def:secres}
A lunar gravity secular resonance occurs whenever there exists an integer vector
$(k_1,k_2,k_3)\in\integer^3\backslash\{0\}$, such that
\beq{secresmoon}
k_1\dot\omega+k_2\dot\Omega+k_3\dot\Omega_M=0\ .
\eeq
We have a solar gravity secular resonance whenever there exist $(k_1,k_2,k_3)\in\integer^3\backslash\{0\}$, such that
\beq{secressun}
k_1\dot\omega+k_2\dot\Omega+k_3\dot\Omega_S=0\ .
\eeq
\end{definition}

We can assume that the rate of variation  $\dot\Omega_S$ is zero, while
for the Moon we will build different models according to which the rate $\dot\Omega_M$
is zero or it is rather equal to  $\dot\Omega_M\simeq -0.053^{\circ}/day$.

As for the debris, we can approximate $\dot \omega$, $\dot \Omega$ by considering only the effect of $J_2$
(\cite{HughesI}):
\beqa{omega12}
\dot\omega &\simeq& 4.98 \Bigl({R_E\over a}\Bigr)^{7\over 2}\ (1-e^2)^{-2}\ (5\cos^2 I-1)\ ^{\circ}/day\ ,\nonumber\\
\dot\Omega &\simeq& -9.97 \Bigl({R_E\over a}\Bigr)^{7\over 2}\ (1-e^2)^{-2}\ \cos I\ ^{\circ}/day\ .
\eeqa
Inserting \equ{omega12} in \equ{secresmoon} or \equ{secressun}, we get an expression which involves the orbital elements
$a$, $e$, $I$, thus providing the location of the secular resonance.

\begin{figure}[h]
\centering
\vglue0.1cm
\hglue0.1cm
\includegraphics[width=6truecm,height=4.5truecm]{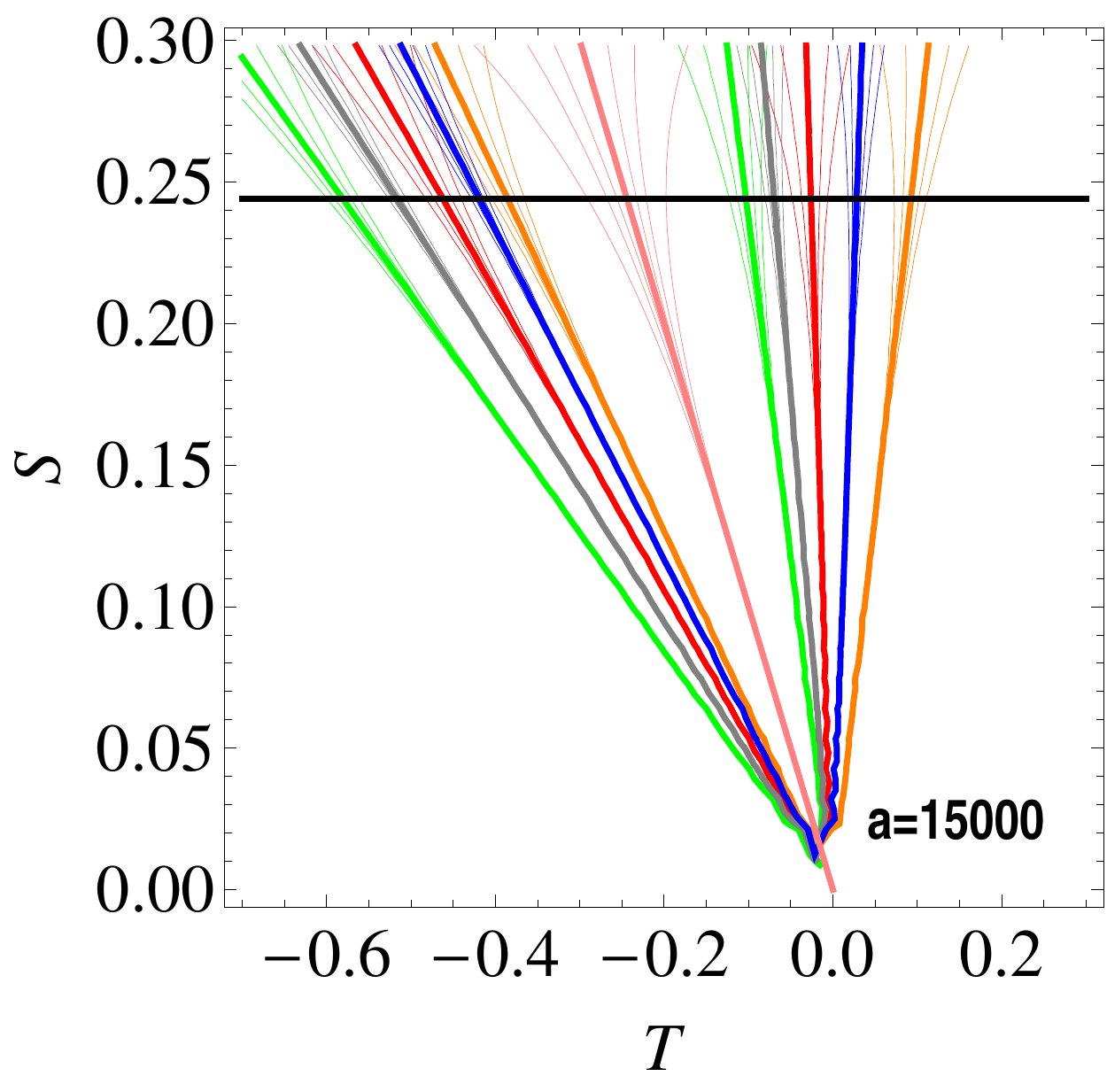}
\includegraphics[width=6truecm,height=4.5truecm]{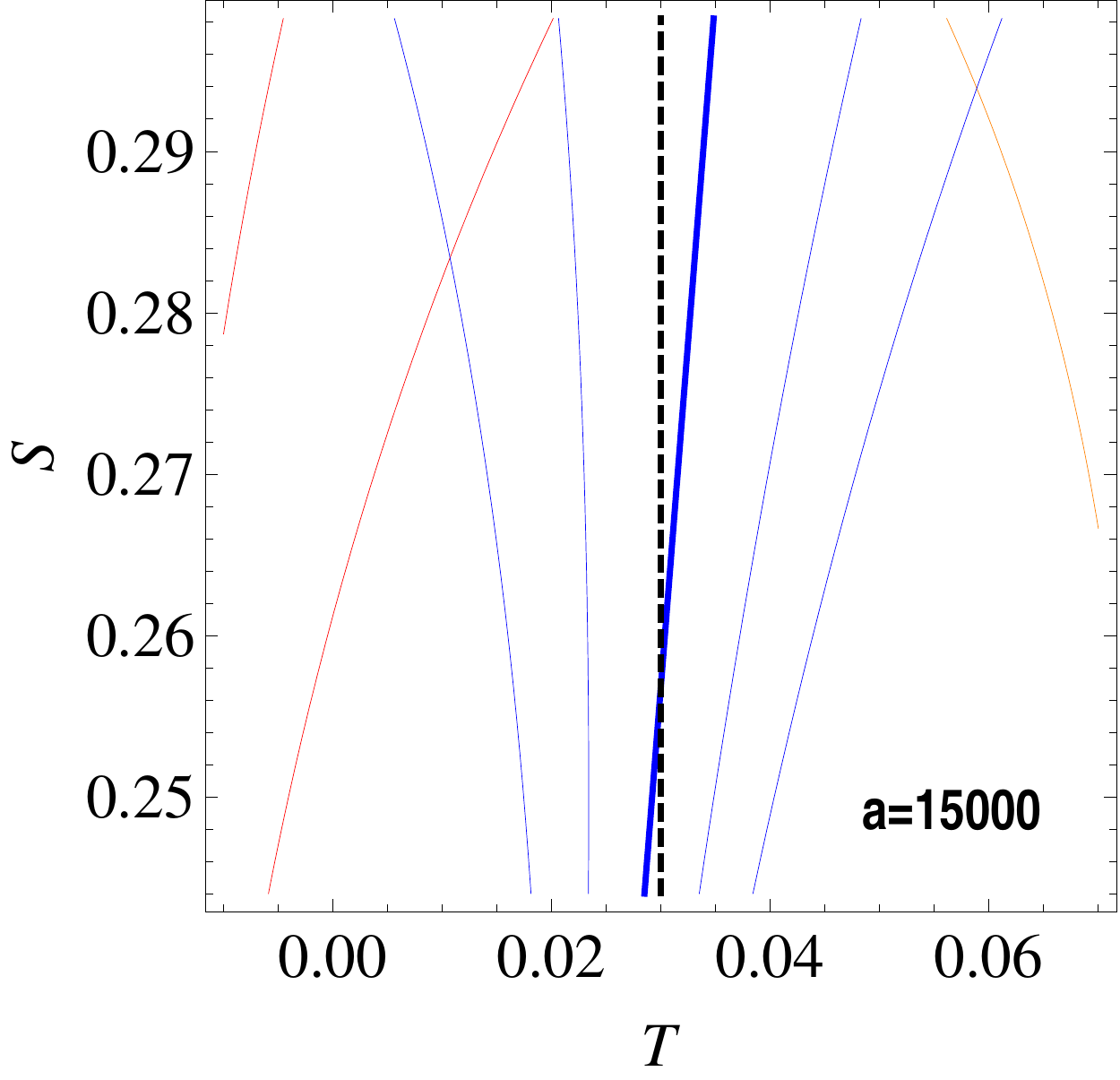}
\includegraphics[width=6truecm,height=4.5truecm]{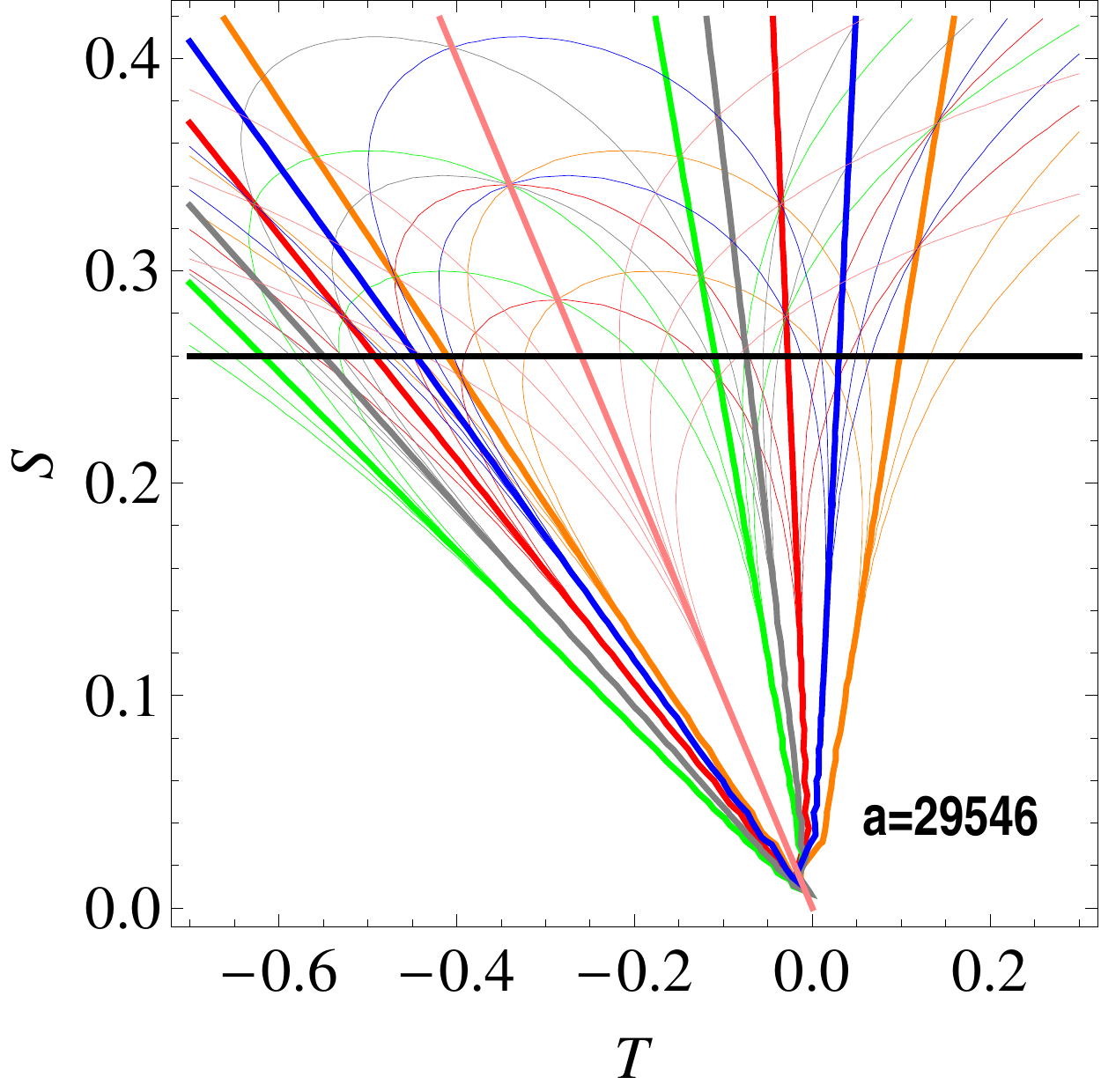}
\includegraphics[width=6.5truecm,height=5truecm]{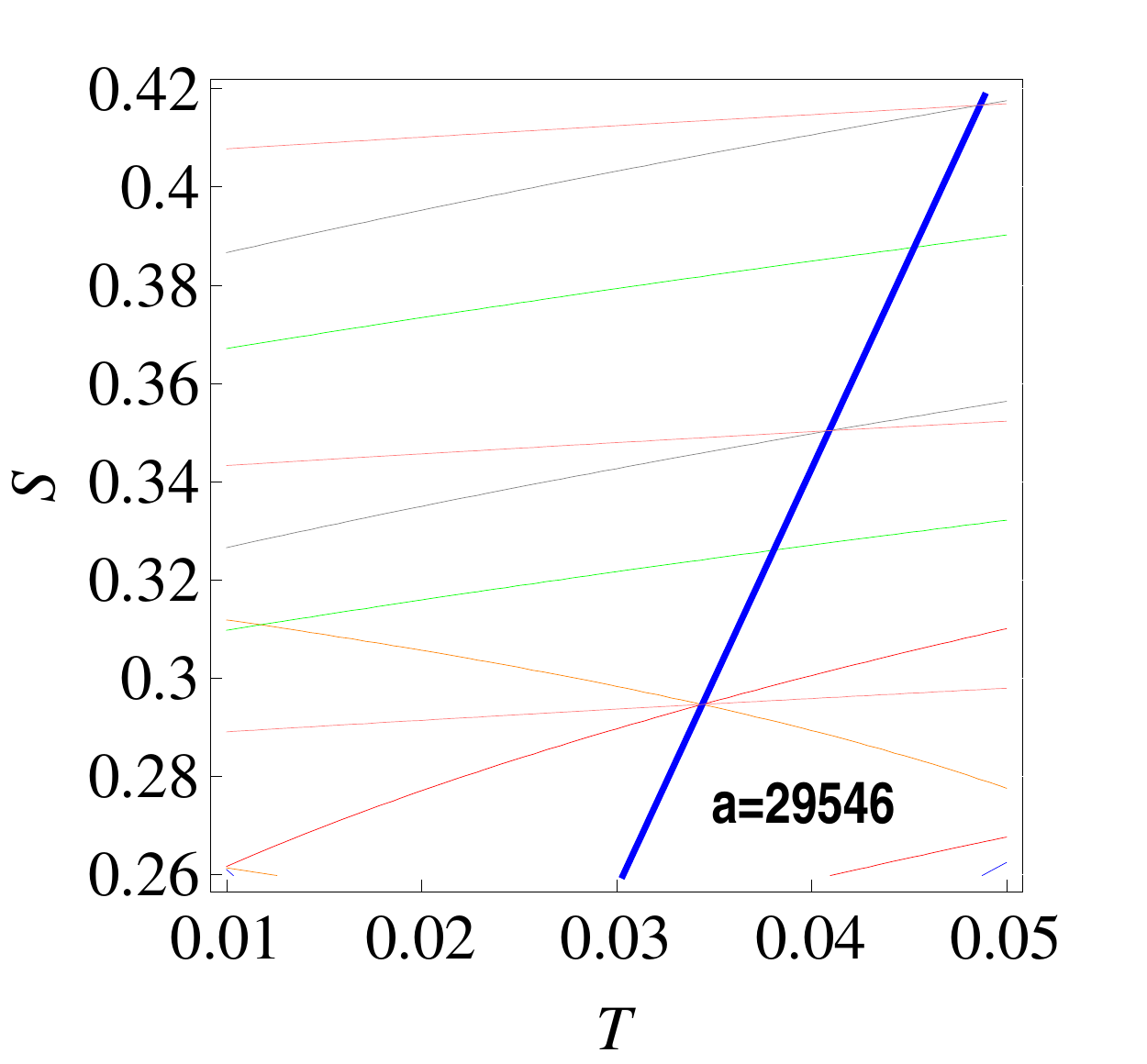}
\vglue0.5cm
\caption{The web structure of resonances in the space of the actions for $a= 15\,000 $ km (upper panels) and  $a= 29\,546 $ km (bottom
panels). The thick curves represent the location of the following {\it exact resonances} (the multiplet component having $s=0$):
$\dot{\Omega}=0$ (pink color, $I=90^\circ$), $\dot{\omega}-\dot{\Omega}=0$ (green color, $I=73.2^\circ$, $I=133.6^\circ$), $2
\dot{\omega}-\dot{\Omega}=0$ (grey color, $I=69.0^\circ$), $I=123.9^\circ$, $\dot{\omega}=0$ (red color, $I=63.4^\circ$, $I=116.6^\circ$),
$2\dot{\omega}+\dot{\Omega}=0$ (blue color, $I=56.1^\circ$, $I=111^\circ$) and $\dot{\omega}+\dot{\Omega}=0$ (orange color, $I=46.4^\circ$,
$I=106.9^\circ$).  The thin curves give the position of the resonances $ (2-2 p) \dot{\omega}+ m \dot{\Omega}+ s \dot{\Omega}_k=0$ with
$p,m=0,1,2$ and $s=-2,-1,1,2$.   The vertical black dashed line (top right panel) corresponds to the values of $T$ used in computing the
Figure~\ref{fig:FLI_model_c_a=15000}. Left panels are obtained for $S\in [0, S_{max}]$, whereas in the right plots $S$ varies from $S_{min}$
to $S_{max}$, as explained in the text.}
\label{WEB_structure}
\end{figure}

A remarkable fact (see \cite{HughesI}) is that some resonances depend
only on the inclination and are independent on $a$, $e$. Precisely, following \cite{HughesI}
we can identify the following classes of lunisolar secular resonances depending only on inclination (see Figure~\ref{WEB_structure}):

\begin{enumerate}
\item[$(i)$]
$\dot\omega=0$, which occurs at the critical inclinations $I=63.4^{\circ}$, $I=116.6^{\circ}$;
\item[$(ii)$]
$\dot\Omega=0$, which corresponds to polar orbits;
\item[$(iii)$]
$\alpha\dot\omega+\beta\dot\Omega=0$ for some nonzero $\alpha$, $\beta\in\integer$.
\end{enumerate}

In this work we are interested to a specific resonance of type $(iii)$ and precisely to the resonance
\beq{res}
2\dot\omega+\dot\Omega=0\ .
\eeq
Using \equ{omega12} and \equ{res}, one can write this resonances as
$$
2\dot\omega+\dot\Omega=\Bigl({R_E\over a} \Bigr)^{7\over 2}\ (1-e^2)^{-2}\ \Big[9.96(5\cos^2 I-1)-9.97\cos I\Big]=0\ ,
$$
whose solutions are $I=56.1^{\circ}$ and $I=111.0^{\circ}$, independently of the values of semimajor axis and eccentricity.

In writing \equ{res} we have implicitly assumed that $\dot\Omega_M=0$ (as we mentioned, the other rates $\dot\omega_M$,
$\dot\omega_S$, $\dot\Omega_S$ can be assumed to be equal to zero). However,
$\Omega_M$ varies periodically and some arguments of $\H_{Moon}$ could depend also on $\Omega_M$. Therefore, besides
$2\dot{\omega}+\Omega=0$, one also has the commensurability relations
\beq{ress}
2 \dot{\omega}+ \Omega+ s\dot{ \Omega}_M=0\ ,\qquad s=-2,-1,1,2\ .
\eeq
This means that the secular resonance splits into a multiplet of resonances. This splitting phenomenon is responsible for the existence of a
very complex web--like background of resonances in the phase space, which leads to a chaotic variation of the orbital elements. An
analytical estimate of the location of the resonance corresponding to each component of the multiplet, as a function of eccentricity and
inclination, can be obtained by using $\eqref{omega12}$ (see, for example, Figure~2 in \citet{ElyHowell} or \citet{aR15}).

To describe properly the dynamics, it is convenient to use resonant variables, which are introduced through the symplectic transformation
$(G,H,\omega, \Omega)\rightarrow (S,T,\sigma, \eta)$ defined by
\begin{equation}
\begin{split}\label{newvariables}
& \sigma=2 \omega+\Omega\ , \qquad S=\frac{G}{2}\ ,\\
& \eta=\Omega\ , \qquad\qquad\ \ T=H-\frac{G}{2}\ .
\end{split}
\end{equation}
Since we expressed the Hamiltonian in Delaunay variables, we represent in Figure~\ref{WEB_structure} the web structure of resonances in the
space of the actions $T$--$S$ introduced in \equ{newvariables}. To avoid confusions that might arise when we speak about a specific
resonance, we will use the terminology {\sl exact resonance} when we refer to the component of the multiplet characterized by $s=0$ in
\equ{ress}, while the expression {\sl whole resonance} means that we refer to all components of the multiplet.

We underline that the units of length and time are normalized so that the geostationary
distance is unity (it amounts to $42\,164.17$ km) and that the period
of the Earth's rotation is equal to $2 \pi$. As a consequence, from
Kepler's third law it follows that $\mu_E=1$. Therefore, unless
the units are explicitly specified, the action variables $L$, $S$ and $T$ are expressed in the above units.

Figure~\ref{WEB_structure} shows the structure of resonances for $a=15\,000$ km (top panels) and $a=29\,546$ km (bottom panels). The colored
curves provide the location of the resonances, while the vertical black dashed line
in the top-right panel is drawn to provide the value of $T$ used in computing the FLI plot for $a=15\,000$ km (see
Figure~\ref{fig:FLI_model_c_a=15000}). In order to show graphical evidence of the splitting phenomenon, Figure~\ref{WEB_structure},  left
panels,  provide the resonant structure for $S\in[0,S_{max}]$, where $S_{max}=\frac{\sqrt{\mu_E a}}{2}$. These plots contain also the
horizontal black line $S=S_{min}$, where $S_{min}$ is computed from the condition that the distance of the perigee cannot be smaller than
the radius of the Earth, that is
$$S_{min}=\frac{1}{2}\, \sqrt{\frac{(2 a- R_E)\mu_E R_E}{a}}\ .$$
Therefore, the interval of interest is $[S_{min}, S_{max}]$. The right panels of Figure~\ref{WEB_structure} magnify the regions associated
to the orbits that do not collide with the Earth (at least for a small interval of time).
Figure~\ref{WEB_structure} shows the complicated interplay of the web of resonances, with multiple crossings
of lines, which correspond to overlapping of resonances, possibly providing a mechanism for the onset of chaos
(\cite{chirikov}, \cite{DRADVR15}).

\section{A comparison of different models}

In order to understand the complicated dynamics of the {\it whole resonance} $2\dot{\omega}+\dot\Omega=0$, we shall simplify further the model
described in the previous Section. In fact, we consider three different models, based on the Hamiltonian function introduced
in \eqref{Hamiltonian}:

 a) The one degree-of-freedom autonomous Hamiltonian, obtained by averaging $\H$ in \eqref{Hamiltonian}
over the \sl fast \rm angle $\eta$ and by neglecting the rates of variation of $\Omega_{M}$.
Indeed, we use the constant value $\Omega_{M}=125.045^\circ$, valid at epoch J2000.

b) The two degrees-of-freedom autonomous Hamiltonian, derived under the assumption that the rate of variation
of $\Omega_M$ is negligible. Again, we use the constant value $\Omega_{M}=125.045^\circ$, valid at epoch J2000.

c) The non--autonomous Hamiltonian $\H$, defined by \eqref{Hamiltonian}.

\vskip.1in

The following sections describe in detail the results which are obtained using models a), b), c).

\subsection{Results for model a)}\label{sec:moda}

\begin{figure}[h]
\centering
\vglue0.1cm
\hglue0.1cm
\includegraphics[width=5.1truecm,height=4truecm]{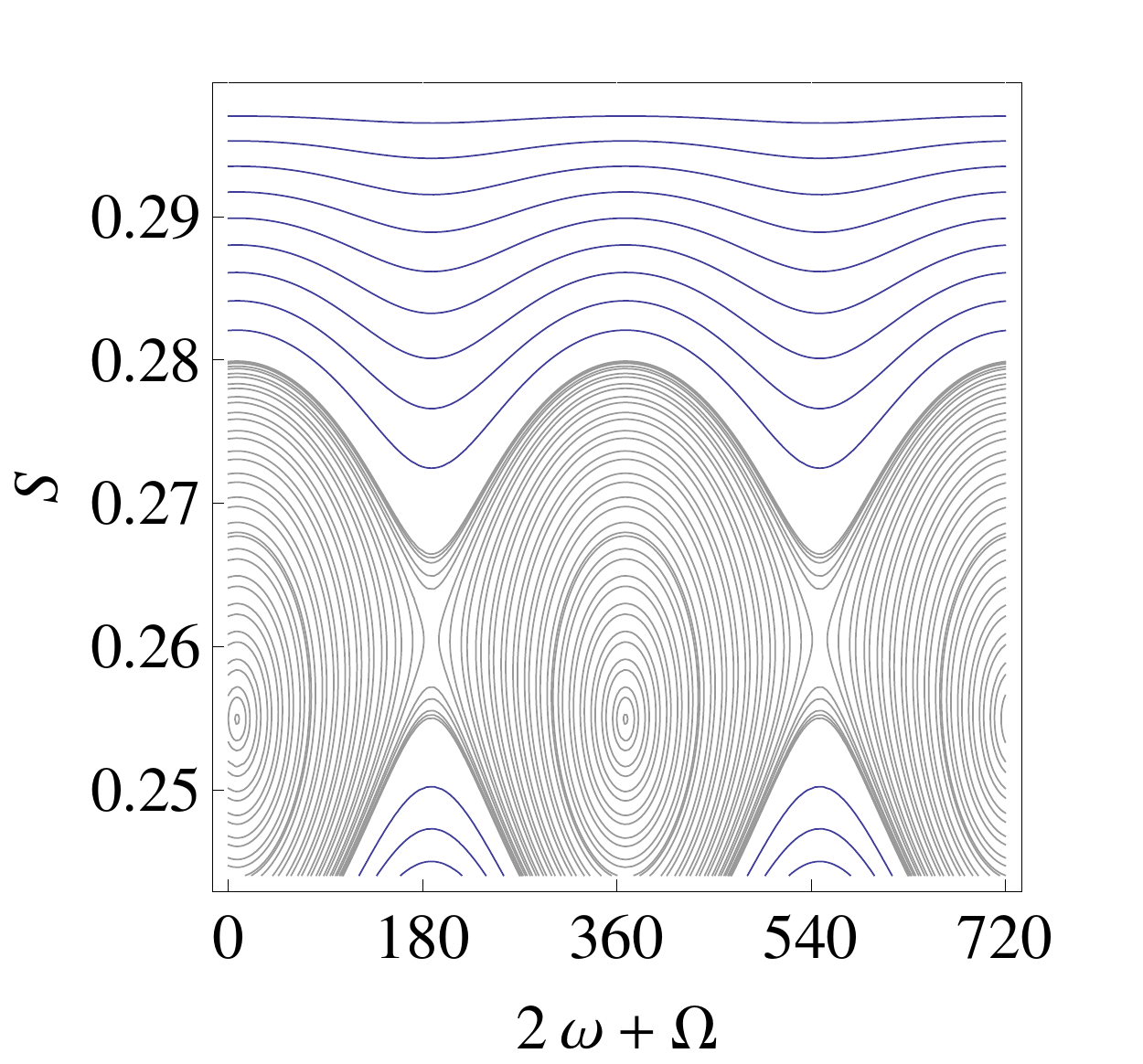}
\includegraphics[width=5.1truecm,height=4truecm]{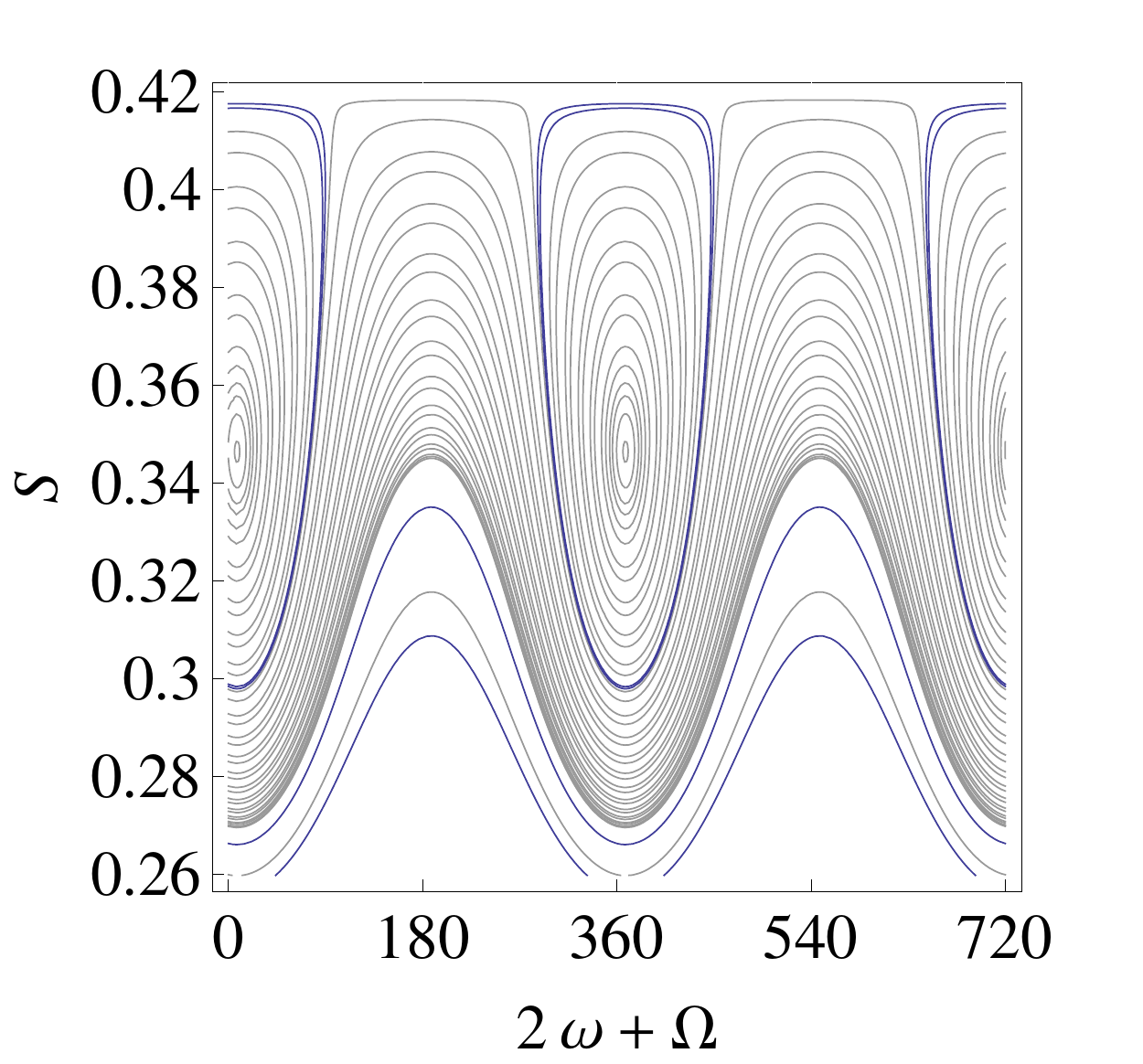}
\includegraphics[width=5.1truecm,height=4truecm]{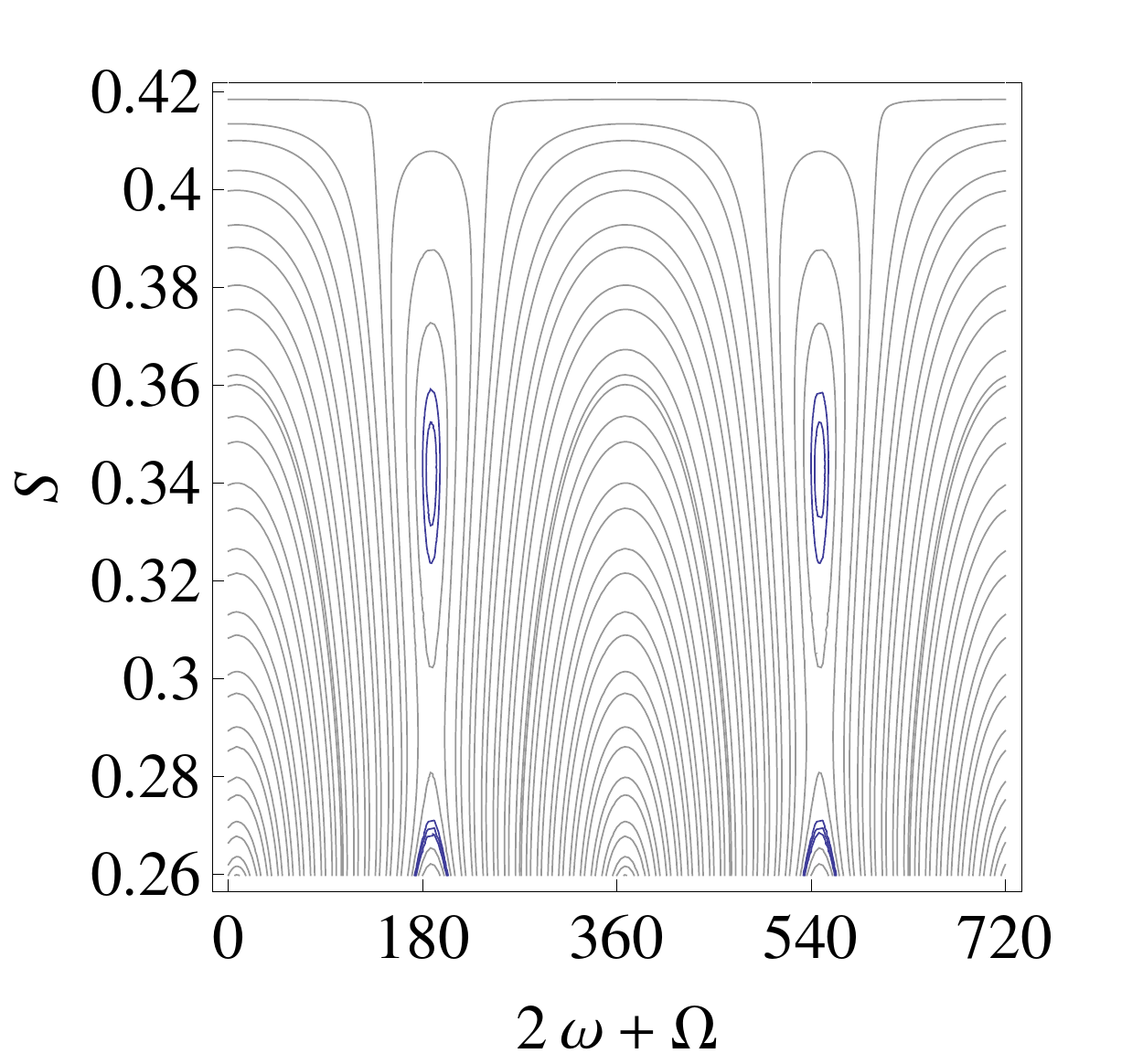}
\vglue0.5cm
\caption{ Phase space portraits for $a=15\,000$ km and $T=0.03$ (left panel), $a=29\,546$ km and $T=0.05$ (middle panel), $a=29\,546$ km and
$T=0.03$ (right panel).}
\label{fig:model_a}
\end{figure}

The results obtained integrating model a), the simplest model as possible,
are shown in Figure~\ref{fig:model_a}, which provides the phase space portraits
for $a=15\,000$ km and $a=29\,546$ km. In order to show more clearly the structure of the phase space, in all figures we represent the
resonant angle $\sigma=2\omega +\Omega$ on intervals longer than $360^\circ$.

Figure~\ref{fig:model_a} shows that for sufficiently small values of the semimajor axis
(left panel) the phase space has a pendulum-like structure, while for larger values of the semimajor axis
(middle and right panels) the pendulum-like model is no longer valid. In fact, for $a=29\,546$ km, a bifurcation phenomenon appears,
showing that there are some cases when a specific resonance cannot be modeled
by a pendulum type system, but one should use a more complex model, referred in the
literature as the \sl extended fundamental model \rm (see \cite{sB01}, \cite{CGPbif} for details).

Comparing the right panel of Figure~\ref{fig:model_a}, obtained for $T=0.03$, with the middle panel of the same Figure~\ref{fig:model_a},
computed for $T=0.05$, we notice the appearance of a new elliptic point, located at $\sigma=180^\circ$. Besides this phenomenon, it is
important to note that the main stable point, which is located at $\sigma=360^\circ$ (or $0^\circ $), changes its position in the action
space as a function of $T$. For instance, for $T=0.05$, this point is located at $S=0.3407$ (or at $e=0.581$, as it follows from
\eqref{Delaunay}, \eqref{newvariables}), while for $T=0.03$, it is positioned at $S=0.26$ (or $e=0.784$). Figure~\ref{fig:model_a} middle
plot reveals the fact that none of the orbits located inside the libration region of the elliptic point will collide with the Earth, while
in Figure~\ref{fig:model_a} right plot, all orbits located inside the libration region associated with the main elliptic point are colliding
orbits.

The integrable model a) gives a clear explanation for the growth of the eccentricity of the satellites and space debris revolving around the
Earth on orbits having an inclination about equal to $56^\circ$. In fact, the growth of the eccentricity is mainly due to the dynamical feature of
the resonance. Inside the libration region, the resonant angle $\sigma=2 \omega + \Omega$ and its conjugated action $S$ vary periodically.
Since,
the eccentricity $e$ is related to $S$ through the relation $e=\sqrt{1-\frac{4S^2}{T^2}}$, then it follows naturally that the eccentricity
varies in time.

\subsection{Results for model b)}\label{sec:modb}

\begin{figure}[h]
\centering
\vglue0.1cm
\hglue0.1cm
\includegraphics[width=5.1truecm,height=4truecm]{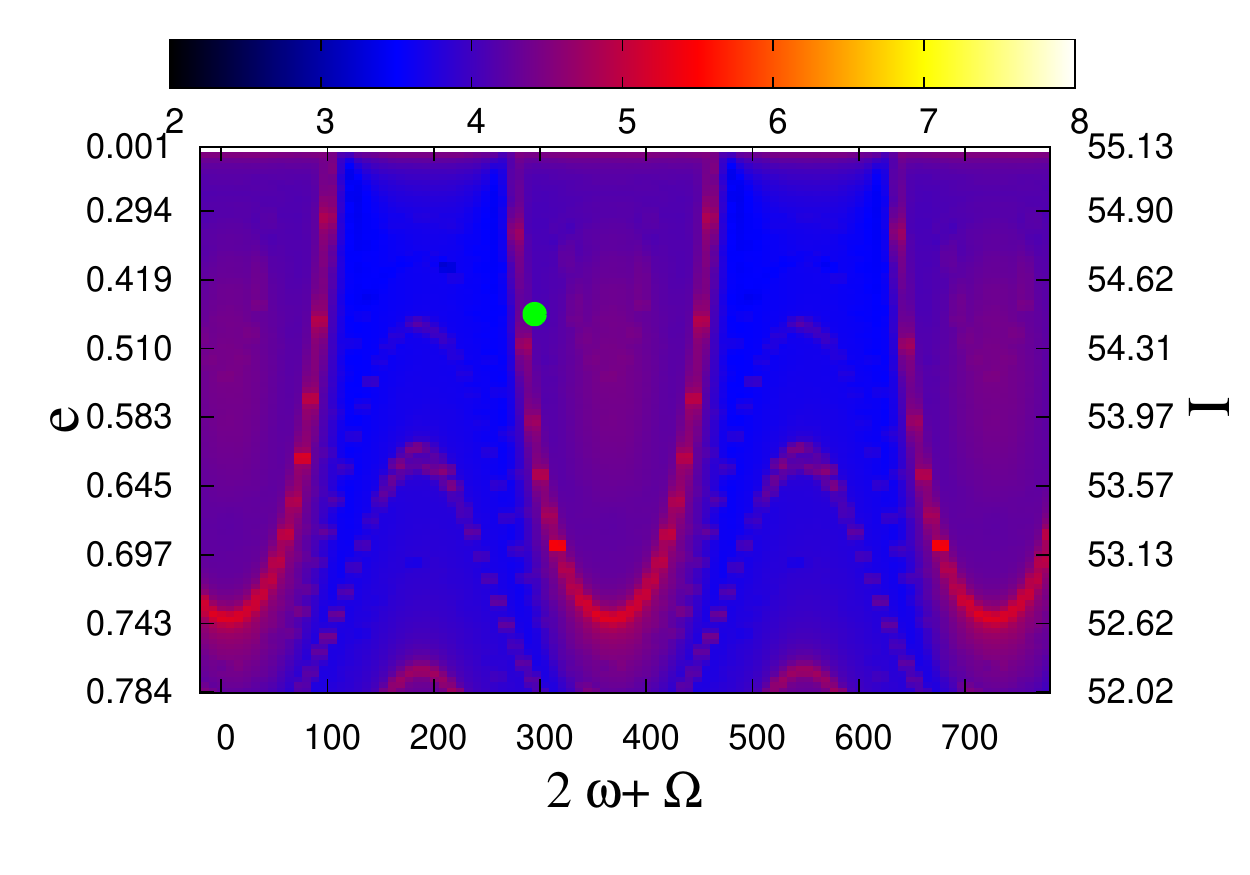}
\includegraphics[width=5.1truecm,height=4truecm]{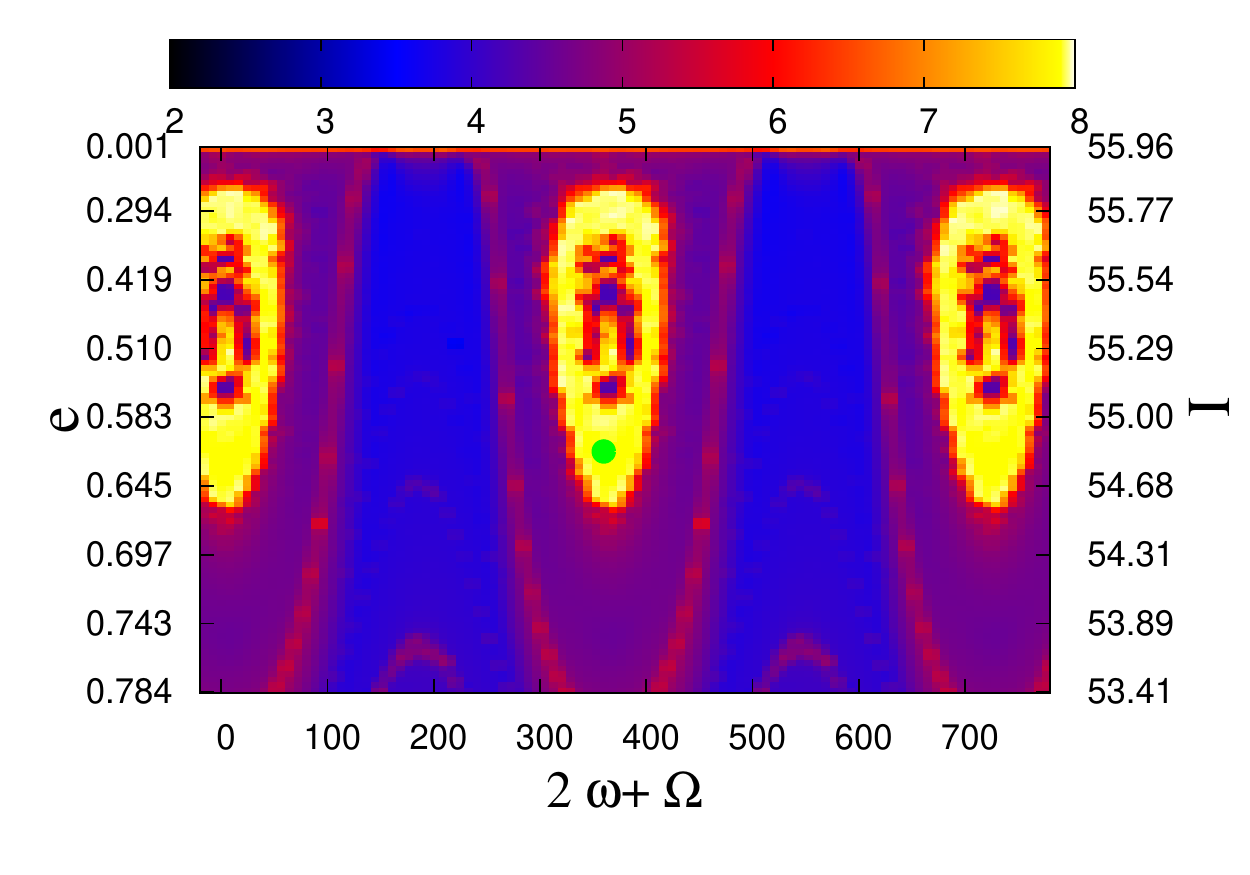}
\includegraphics[width=5.1truecm,height=4truecm]{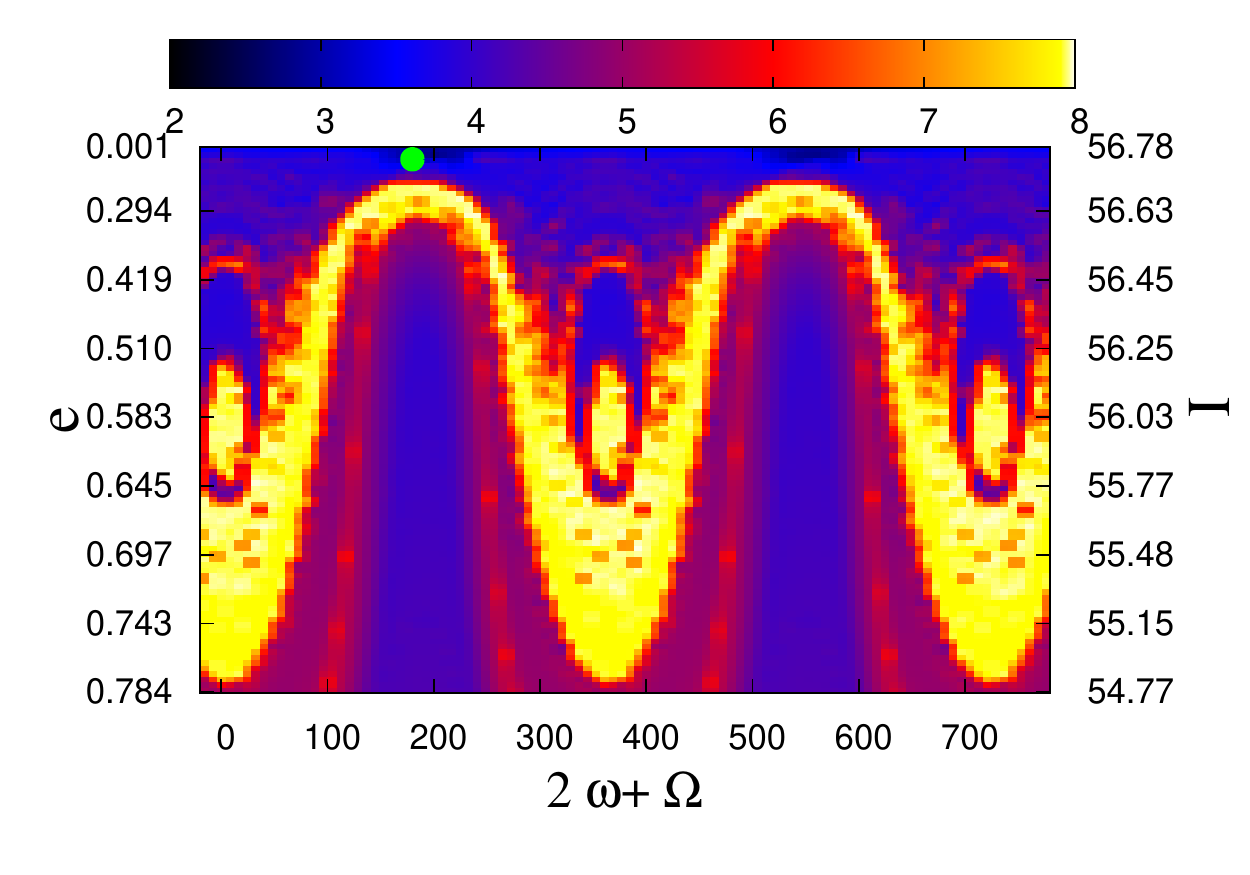}
\vglue0.5cm
\caption{FLIs for the model b), for $a=29\,546$ km, $\Omega=180^\circ$ and: $T=0.06$ (left), $T=0.05$ (middle), $T=0.04$ (right).
Each plot contains one green circle. These circles represent the orbits analyzed in
Figure~\ref{fig:orbits_T=0_06_T=0_05_T=0_04_Om=180_model_b}.}
\label{fig:FLI_model_b_Om=180}
\end{figure}

\begin{figure}[h]
\centering
\vglue0.1cm
\hglue0.1cm
\includegraphics[width=5truecm,height=4truecm]{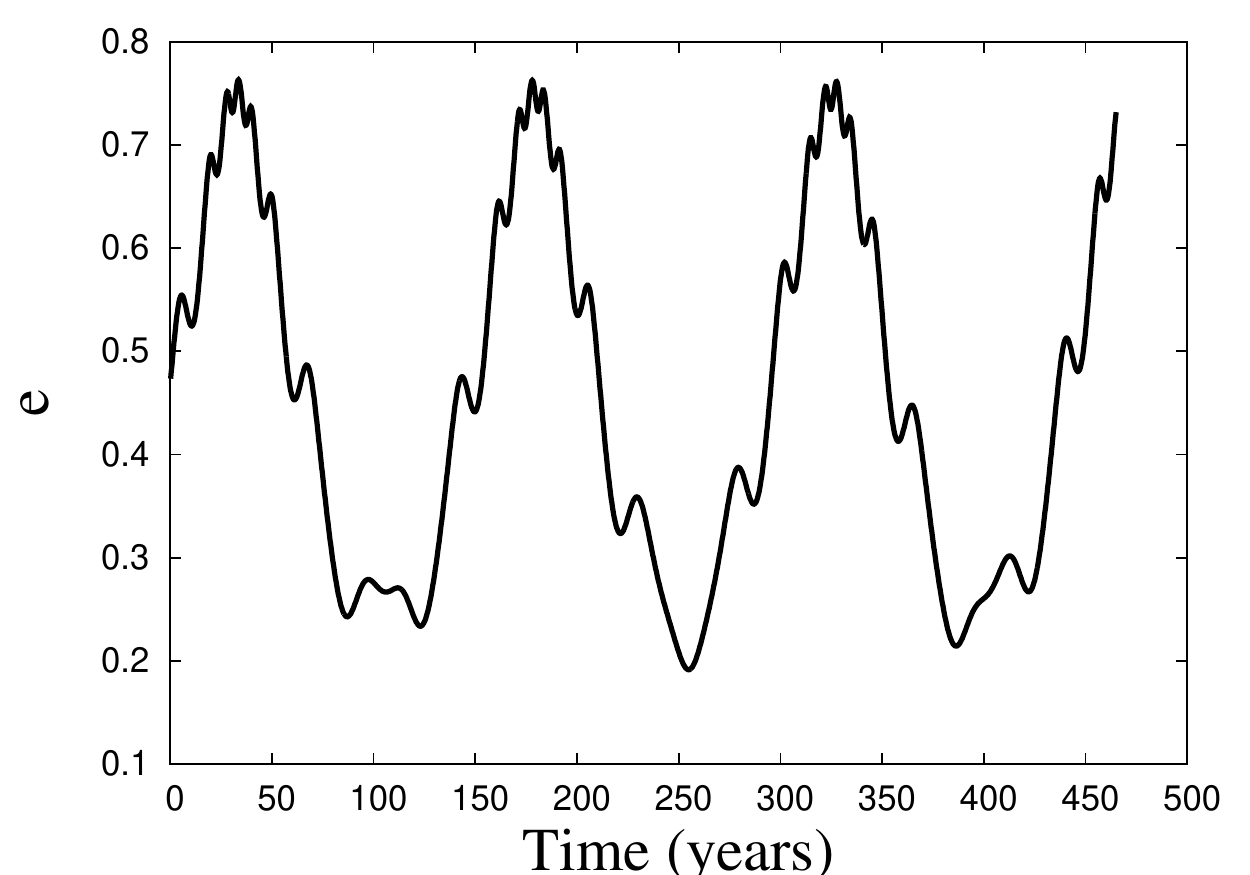}
\includegraphics[width=5truecm,height=4truecm]{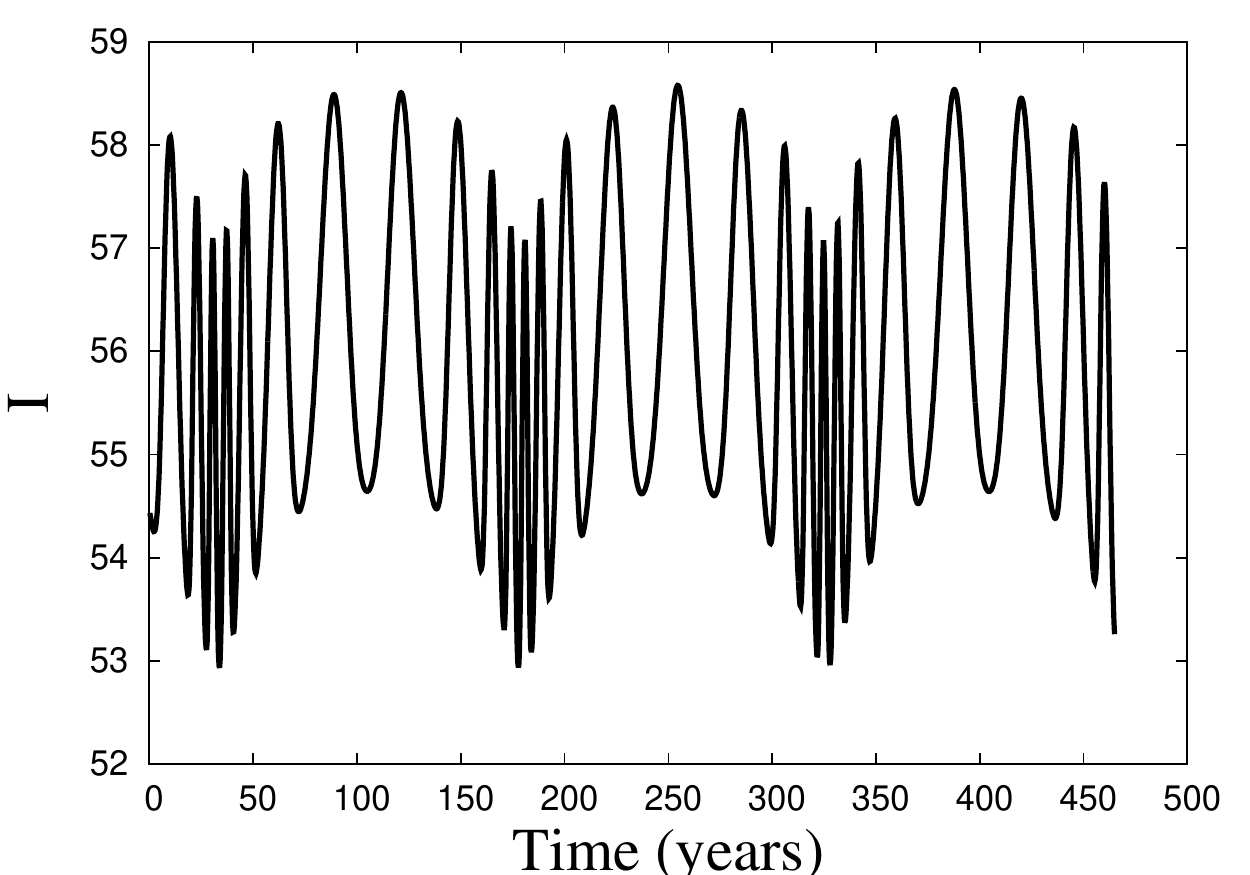}
\includegraphics[width=5truecm,height=4truecm]{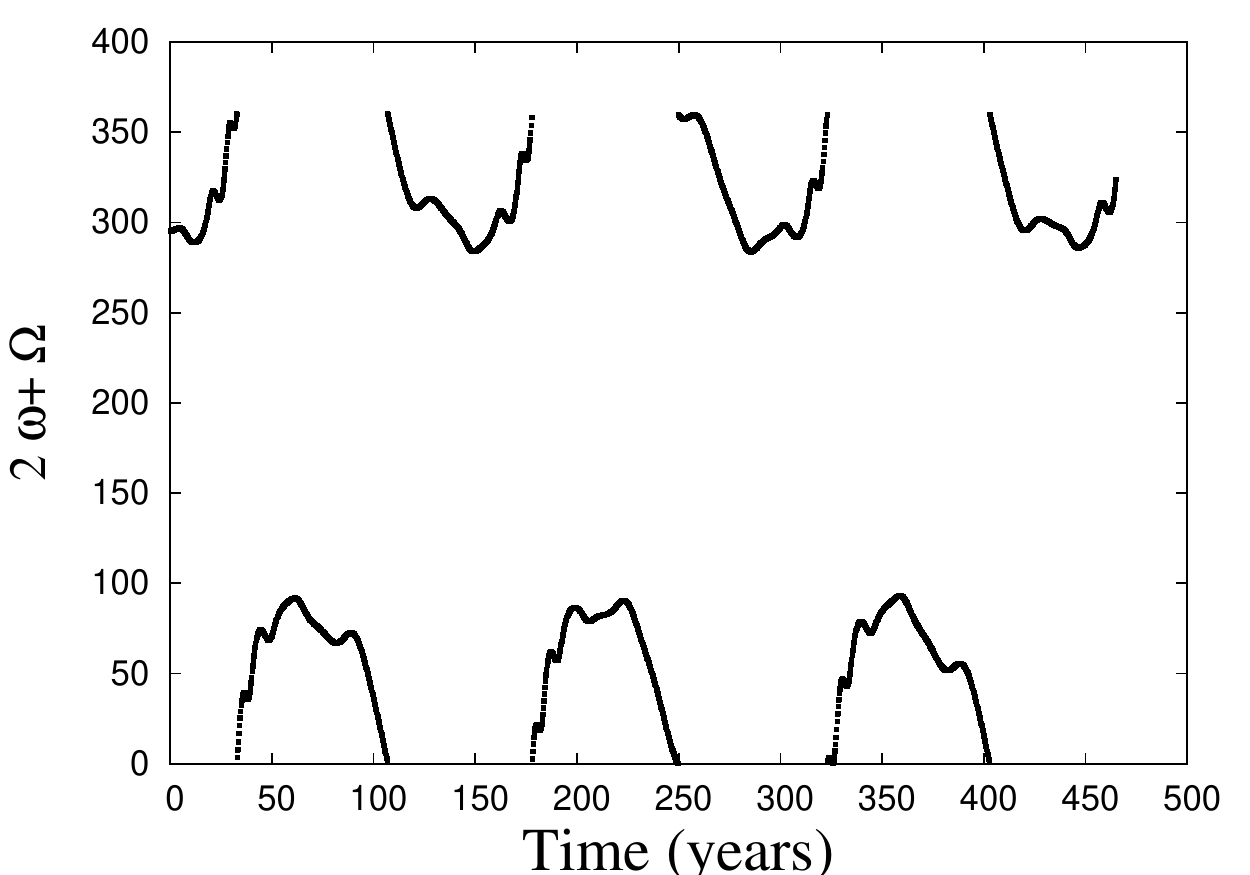}
\includegraphics[width=5truecm,height=4truecm]{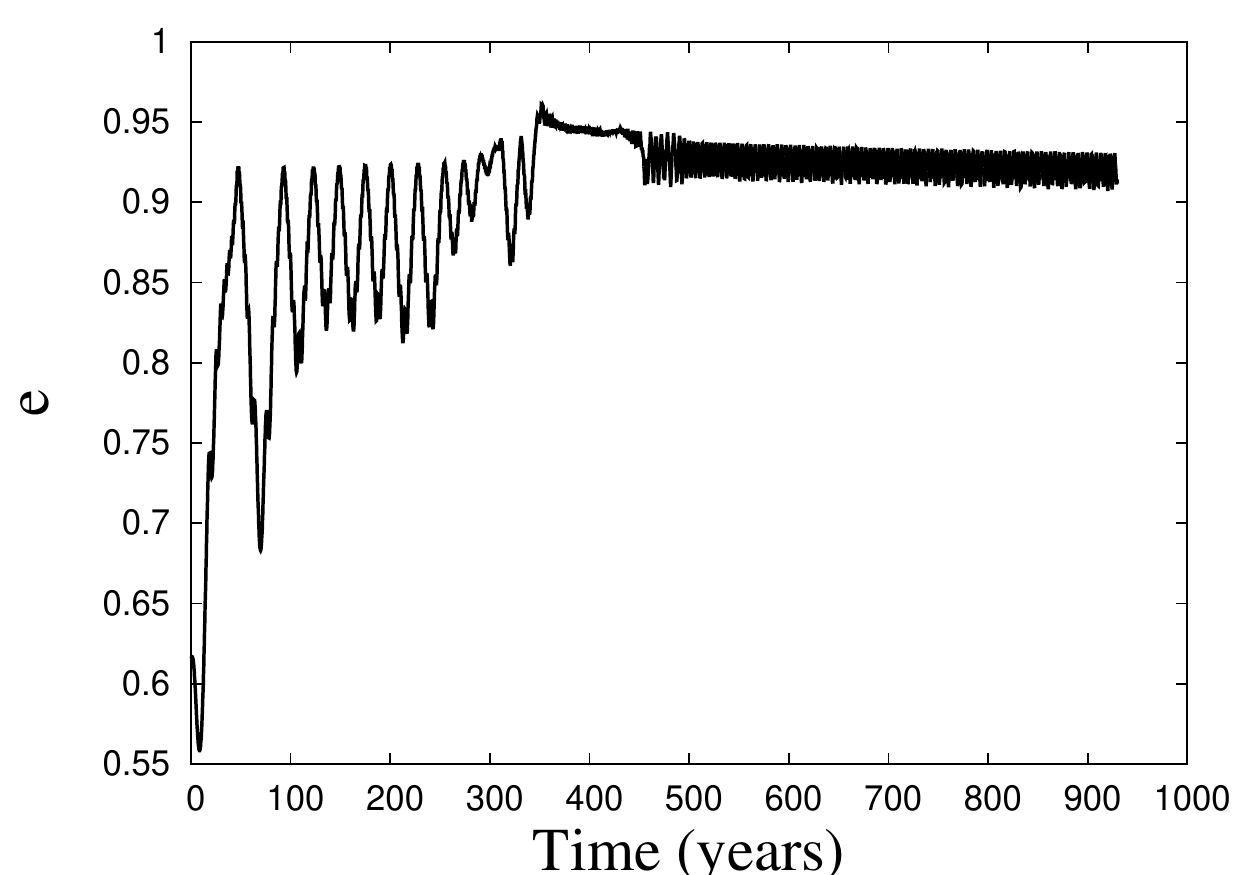}
\includegraphics[width=5truecm,height=4truecm]{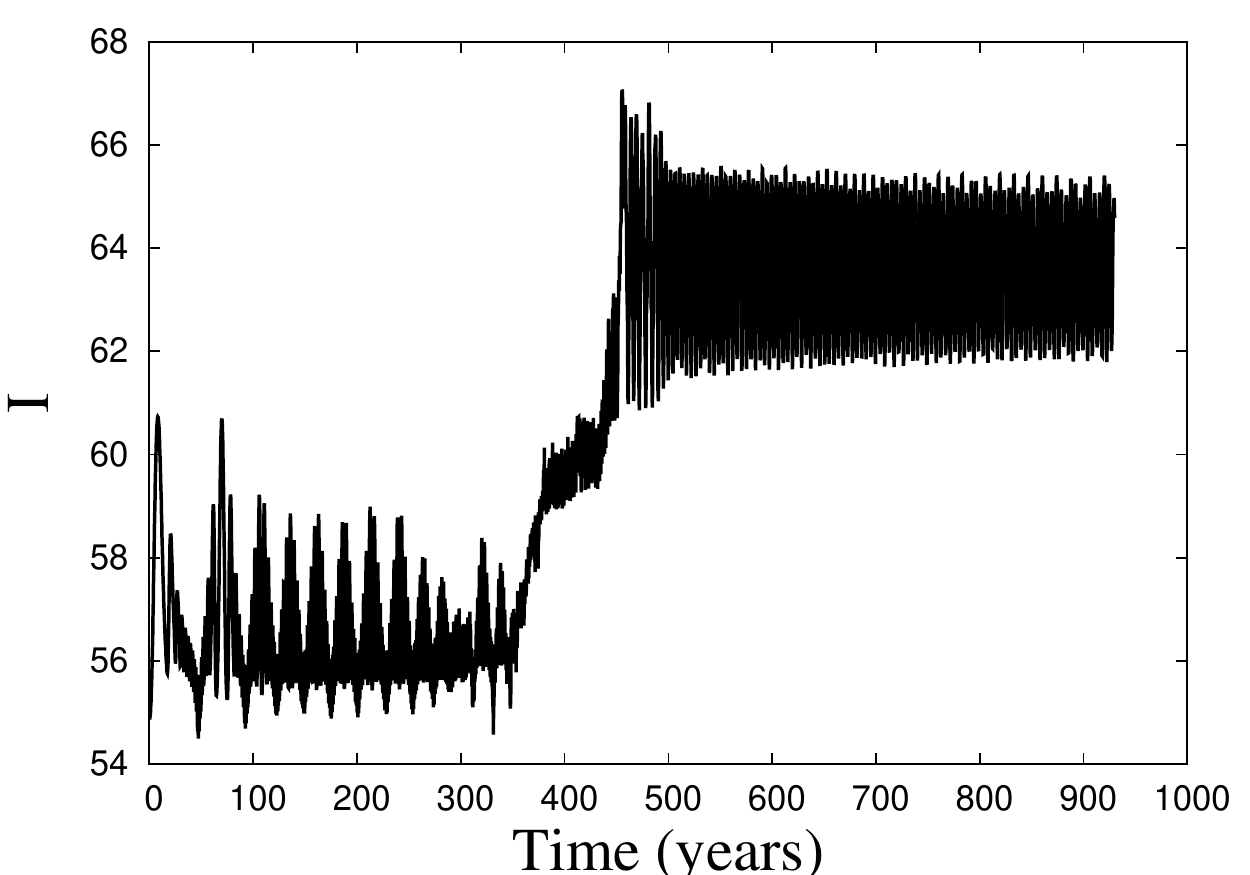}
\includegraphics[width=5truecm,height=4truecm]{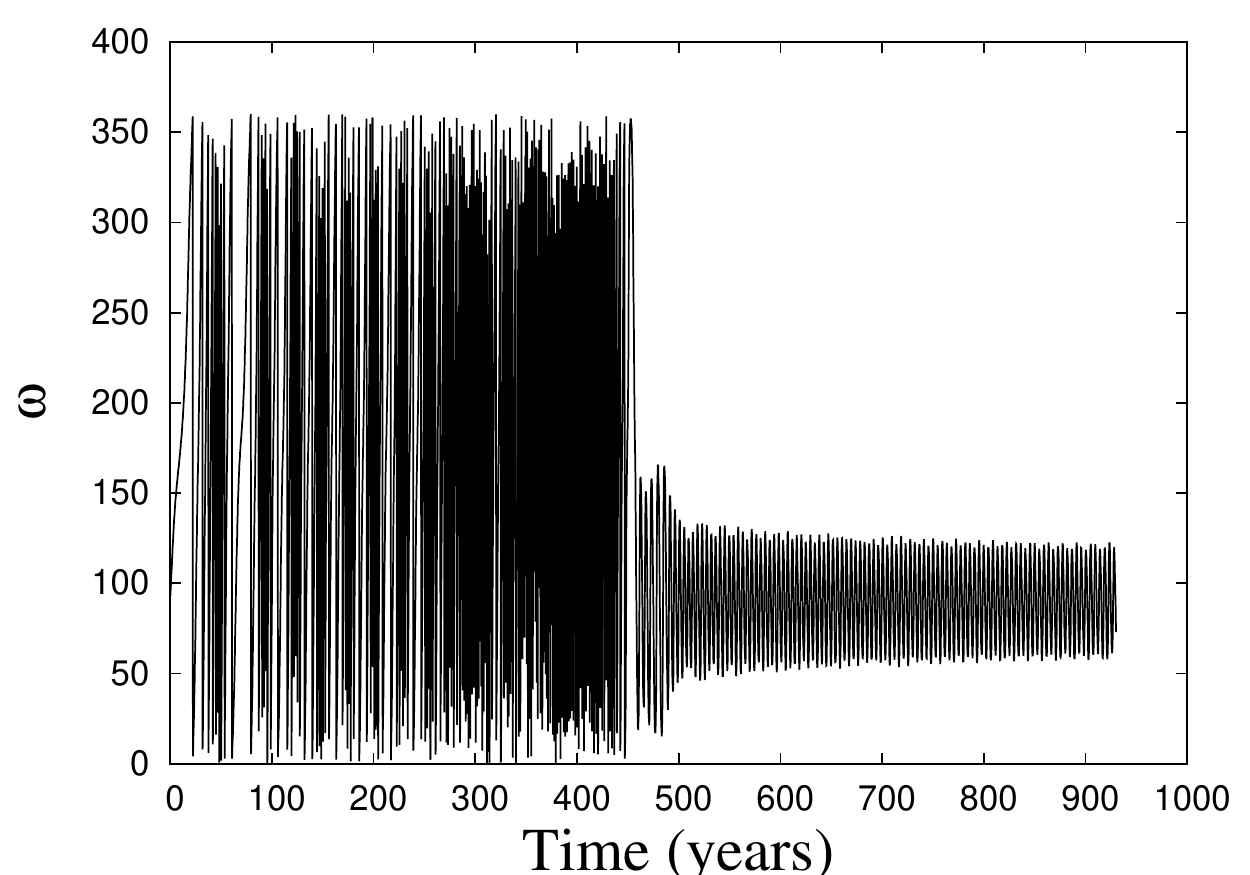}
\includegraphics[width=5truecm,height=4truecm]{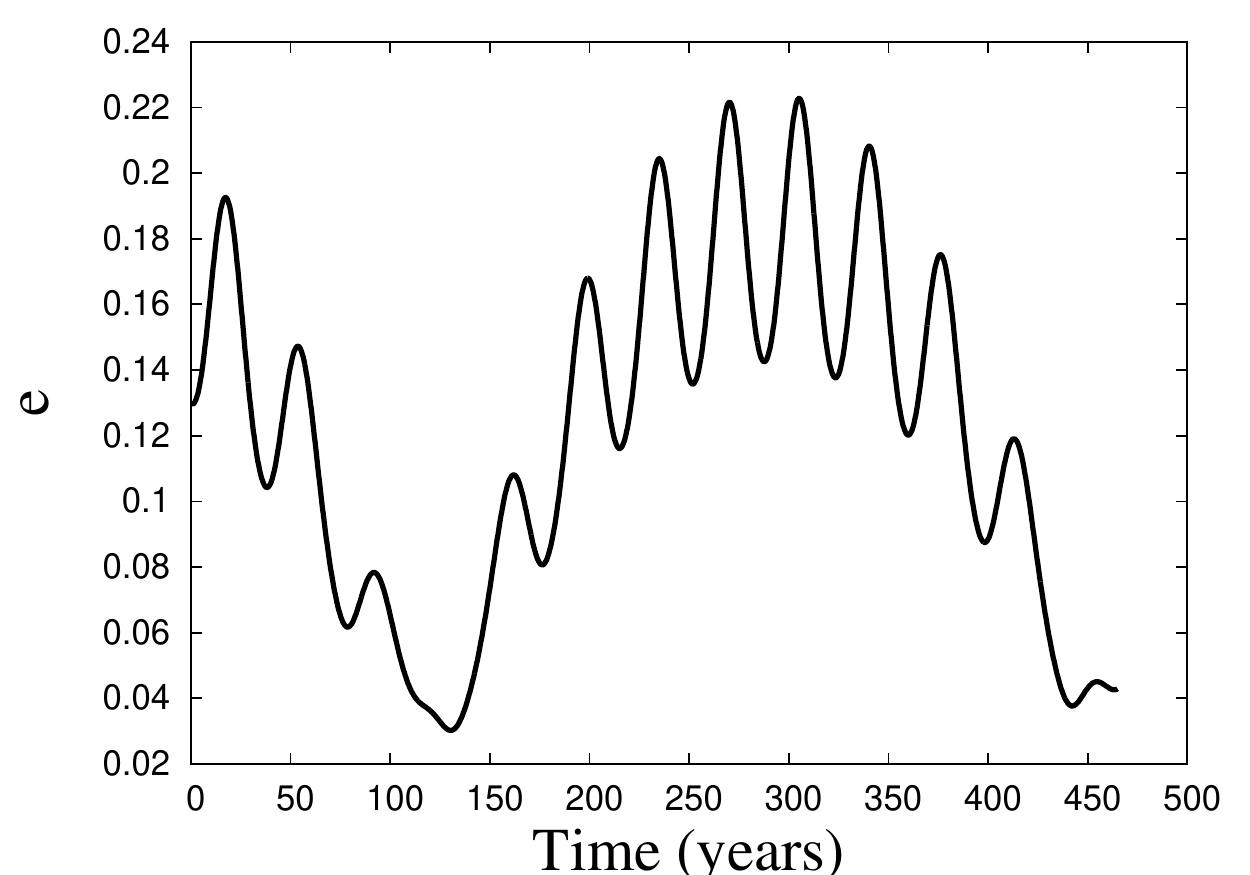}
\includegraphics[width=5truecm,height=4truecm]{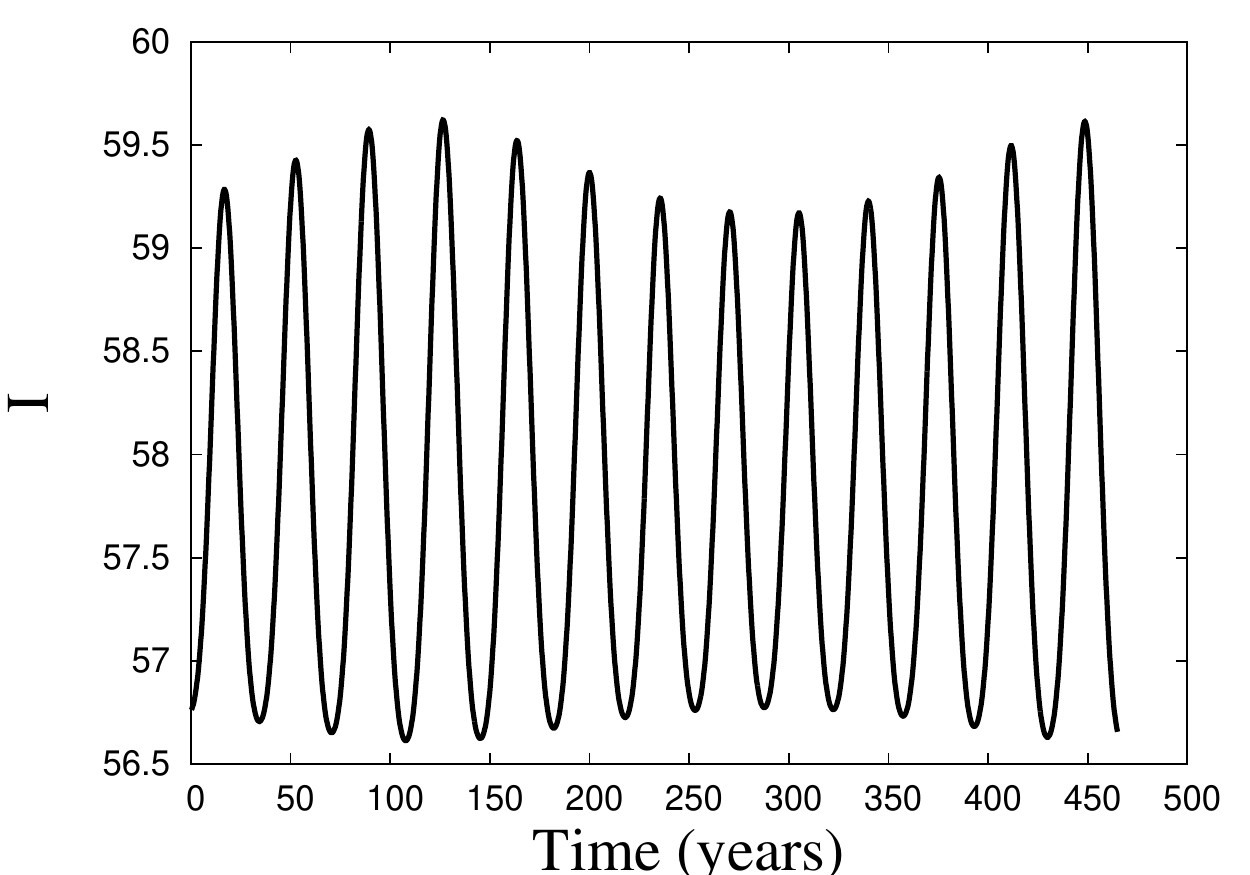}
\includegraphics[width=5truecm,height=4truecm]{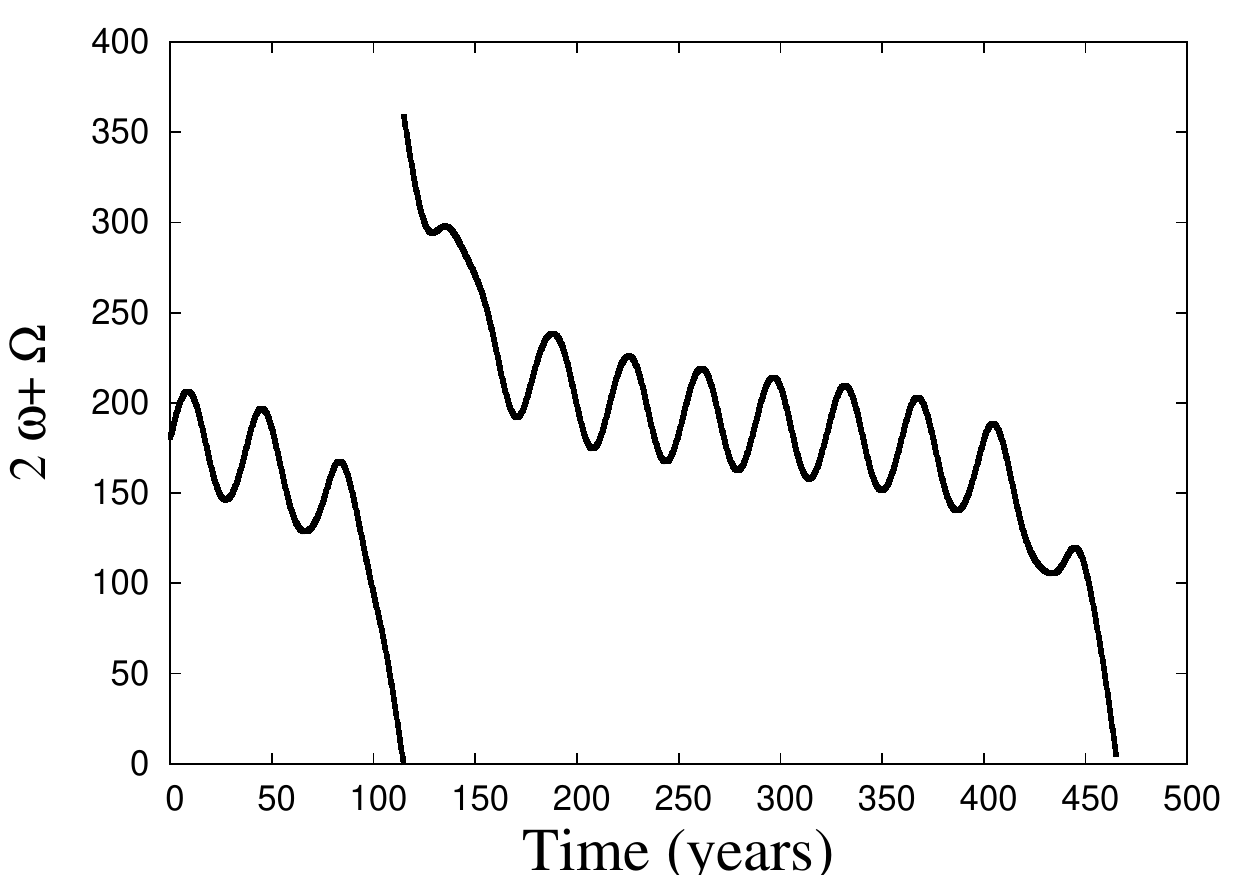}
\vglue0.5cm
\caption{Integration of the orbits having the initial conditions $\Omega=180^\circ$ and: $\sigma=295^\circ$, $T=0.06$, $S=0.37$ (or
$e=0.467$, $I=54.47^\circ$) (top plots); $\sigma=360^\circ$, $T=0.05$, $S=0.33$ (or $e=0.615$, $I=54.85^\circ$) (middle plots);
$\sigma=180^\circ$, $T=0.04$, $S=0.415$ (or $e=0.13$, $I=56.76^\circ$) (bottom plots). }
\label{fig:orbits_T=0_06_T=0_05_T=0_04_Om=180_model_b}
\end{figure}

\begin{figure}[h]
\centering
\vglue0.1cm
\hglue0.1cm
\includegraphics[width=5truecm,height=4truecm]{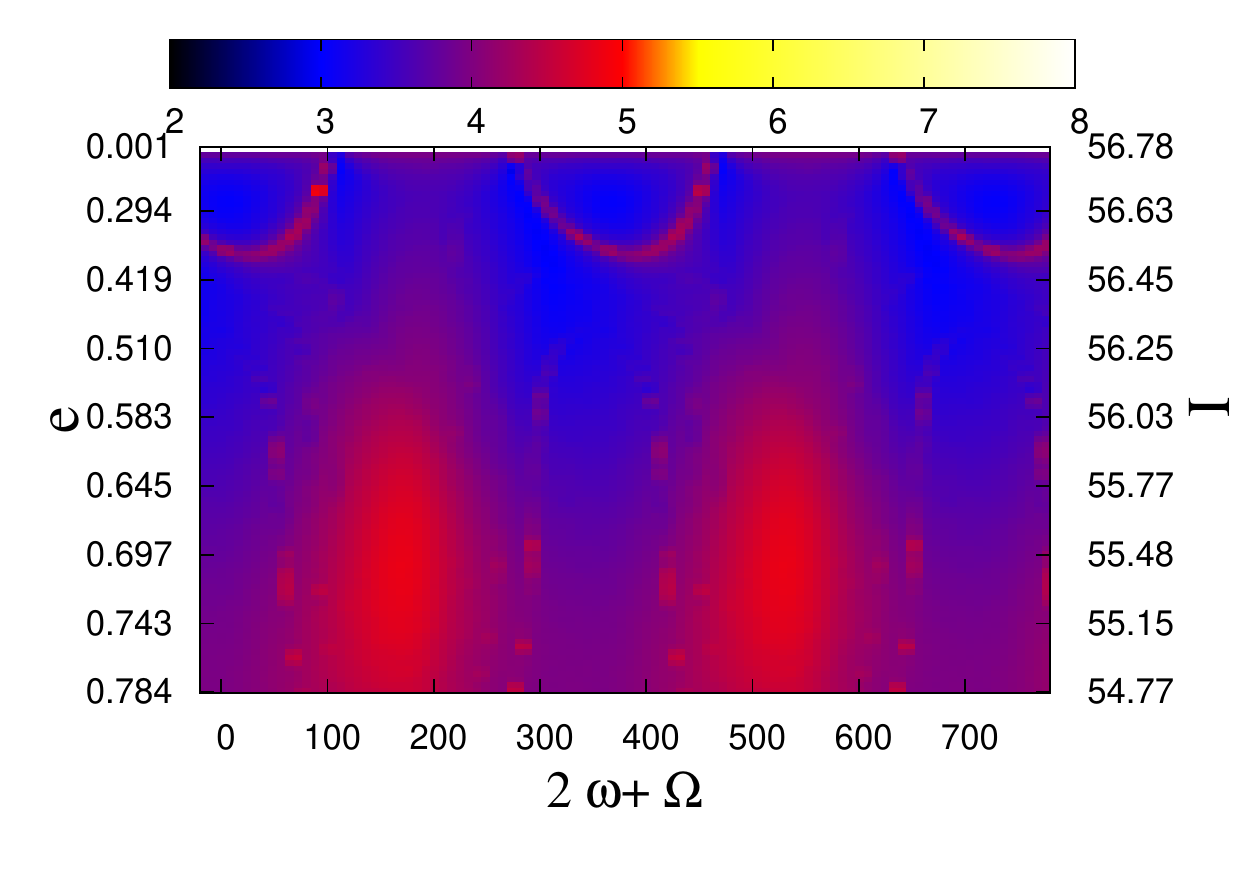}
\includegraphics[width=5truecm,height=4truecm]{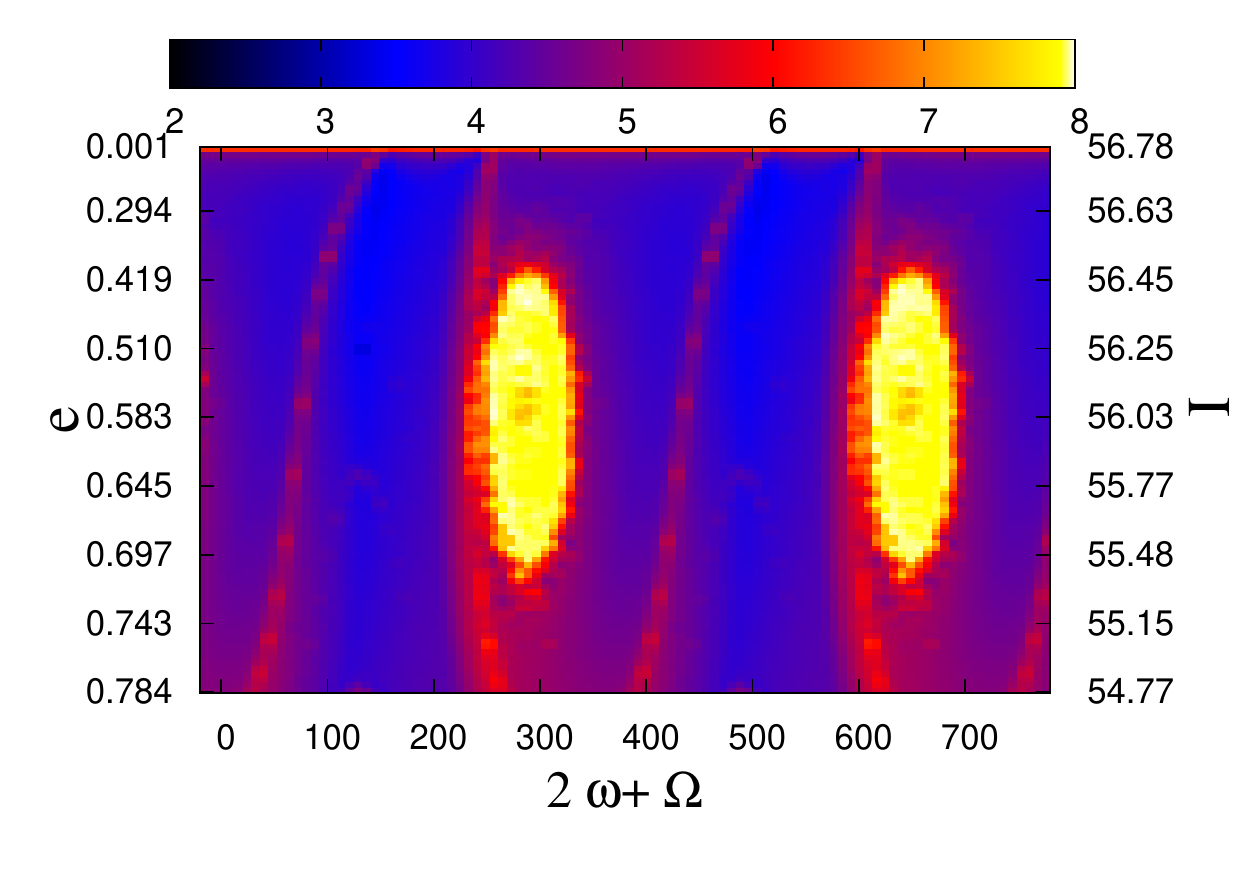}
\includegraphics[width=5truecm,height=4truecm]{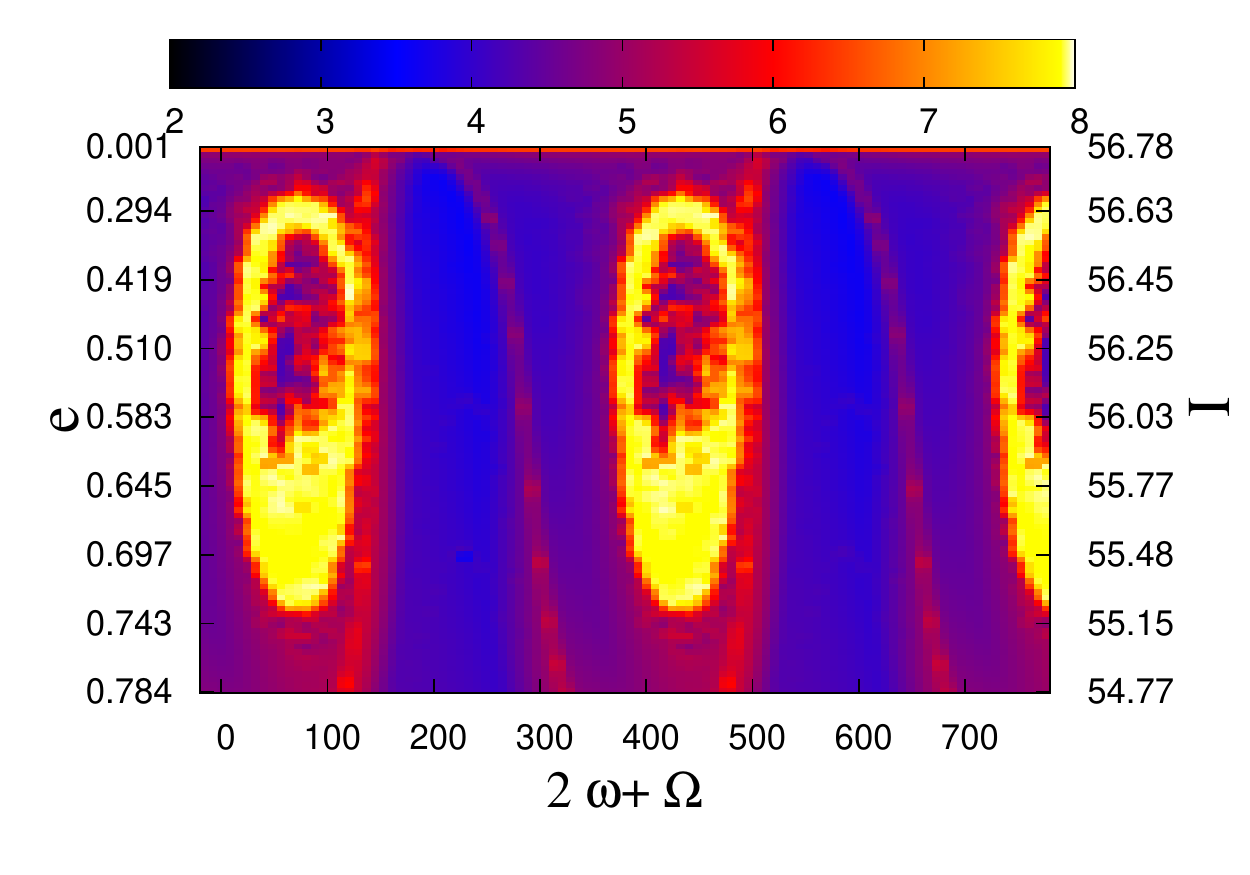}
\vglue0.5cm
\caption{FLIs for the model b), for $a=29\,546$ km, $T=0.04$ and: $\Omega=0^\circ$ (left); $\Omega=90^\circ$ (middle); $\Omega=270^\circ$
(right).}
\label{fig:FLI_model_b_Om=0_90_270}
\end{figure}

To analyze model b) we us the Fast Lyapunov Indicators (hereafter, FLI), which are defined as the largest Lyapunov characteristic exponents at a fixed time (compare with \cite{CGmajor}).
We provide the definition of FLI in Appendix~\ref{app:FLI}.
Their values provide a numerical indication of
the stable (low values) and chaotic (high values) behavior of the dynamical system as the initial conditions or
some internal parameters are varied.

We shall focus on $a=29\,546$ km, because for $a=15\,000$ km the phase plane $ \sigma$--$S$, even in the case of the full model c), is
similar to a pendulum, as it is shown in Figure~\ref{fig:FLI_model_c_a=15000}.

The results for model b) are given in Figures~\ref{fig:FLI_model_b_Om=180}--\ref{fig:FLI_model_b_Om=0_90_270}. Thus, given $a=29\,546$ km
and a value for $T$, we compute a grid of $100 \times 100 $ points of
the $\sigma$--$S$ plane, where the resonant angle ranges in the interval $[0^\circ, 360^\circ]$ (also here we use a larger interval just to show
better the structure of the phase space), while $S$ spans the interval $[S_{min}, S_{max}]$. However, instead of displaying $S$ on the
vertical axis, in each plot we show the eccentricity values (on the left) and the inclination values (on the right), computed by using the
relations \eqref{Delaunay} and \eqref{newvariables} for given values of $T$.
In all plots that represent the FLI values, we use the ranges corresponding to those used in the right panels of Figure~\ref{WEB_structure}.
The relation among $S$, $T$, $e$ and $I$ is trivial; for instance, the value $e=0.784$ from the left panel of
Figure~\ref{fig:FLI_model_b_Om=180} corresponds to the value $S=0.26$ from the top right panel of Figure~\ref{WEB_structure}, while the
value $I=52.02^\circ$ from the same left panel of Figure~\ref{fig:FLI_model_b_Om=180} corresponds to the values $S=0.26$ and $T=0.06$.

Although the initial conditions are set such that the initial orbits have the perigee larger than $R_E$, since we are interested in
understanding the mean dynamical features of the $2 \dot{\omega}+\dot{\Omega}=0$ resonance, during the total time of integration, we neglect
the Earth's dimensions. Namely, we propagate each orbit up to $465$ years (equal to $25 \times 18.6$ years), even if at some intermediate
time the perigee distance becomes smaller than the radius of the Earth.

As we mentioned in Section~\ref{sec:moda}, for large values of the semimajor axis in model a), the phase space is much more complicated than
the one associated to the pendulum model. The complexity increases when we consider the two degrees-of-freedom autonomous Hamiltonian
of model b). In fact, the manifolds defined by $\H(S,T,\sigma,\eta)=const.$ have dimension three in the four dimensional phase space
$\mathbb{R}^2\times \mathbb{T}^2$. This makes difficult the visualization of phase portraits or even the interpretation
of the FLI plots. However, we can draw some conclusions from
Figures~\ref{fig:FLI_model_b_Om=180} and \ref{fig:FLI_model_b_Om=0_90_270}, obtained by projecting the phase space
on the plane $(\sigma, S)$, for fixed values of $T$ and $\eta$.

In fact, we underline three aspects concerning the global dynamics, which are revealed by the model b), namely:  the amplitude of resonance
depends on the values of both canonical variables $T$ and $\eta$. For some values of the canonical variables, the resonances
$2\dot{\omega}+\dot{\Omega}=0$ and $\dot{\omega}=0$ overlap; the bifurcation phenomenon, revealed by the model a), is observable both in
this case but also in the case of the full model c).

The plots shown in Figure~\ref{fig:FLI_model_b_Om=180} are obtained for $\eta=180^\circ$ and different values of its conjugated action $T$,
while Figure~\ref{fig:FLI_model_b_Om=0_90_270} shows some results obtained for the same value of $T$ and various values of $\eta$. Moreover,
in order to have a clear idea about the patterns shown in these plots, in Figure~\ref{fig:orbits_T=0_06_T=0_05_T=0_04_Om=180_model_b} we
represent the evolution of the eccentricity, inclination and the resonant angle for three distinct orbits. Thus, the orbit depicted by the
top plots of Figure~\ref{fig:orbits_T=0_06_T=0_05_T=0_04_Om=180_model_b} (the green circle in the left panel of
Figure~\ref{fig:FLI_model_b_Om=180}) is located inside the libration region; the eccentricity and resonant angle vary periodically. In the
middle panels of Figure~\ref{fig:orbits_T=0_06_T=0_05_T=0_04_Om=180_model_b} (see also the green circle of the middle panel of
Figure~\ref{fig:FLI_model_b_Om=180}) we consider an orbit located inside the region where the resonances $2
\dot{\omega}+\dot{\Omega}=0$ and $\dot{\omega}=0$
are so close that there is a non negligible interaction; we integrate the orbit over a longer time (930 years), even if it is a colliding
orbit just to show the strong interaction of the above mentioned resonances. Over a period of 350 years the orbit is located inside the
libration region of the resonance  $2\dot{\omega}+\dot{\Omega}=0$, then, after an interval of time, it
escapes from that resonance and it is rather
captured into the critical
inclination resonance. Finally, the bottom plots of Figure~\ref{fig:orbits_T=0_06_T=0_05_T=0_04_Om=180_model_b} correspond also to a
resonant orbit (the green circle of the right panel of Figure~\ref{fig:orbits_T=0_06_T=0_05_T=0_04_Om=180_model_b}):
they do not belong to the main
resonant libration region, but rather to the resonant small region which appears as a result of the bifurcation phenomenon,
already described by the model a).

In conclusion, the global dynamics revealed by the model b) is very complex: overlapping of resonances (the yellow regions\footnote{ For the critical inclination resonance, the stable
equilibrium points are located at $\omega=90^\circ$ and $\omega=270^\circ$.} in
Figures~\ref{fig:FLI_model_b_Om=180}, \ref{fig:FLI_model_b_Om=0_90_270}), bifurcations and, as a consequence, the existence of equilibria
at large eccentricities as well as at small eccentricities, variation of the amplitude of the resonance as a function of $T$ and $\eta$
(compare, for instance, the small libration zone of the left plot of Figure~\ref{fig:FLI_model_b_Om=0_90_270} with the large libration
regions from the middle and right plots again of Figure~\ref{fig:FLI_model_b_Om=0_90_270}).

\subsection{Results for model c)}\label{sec:modc}

\begin{figure}[h]
\centering
\vglue0.1cm
\hglue0.1cm
\includegraphics[width=5truecm,height=4truecm]{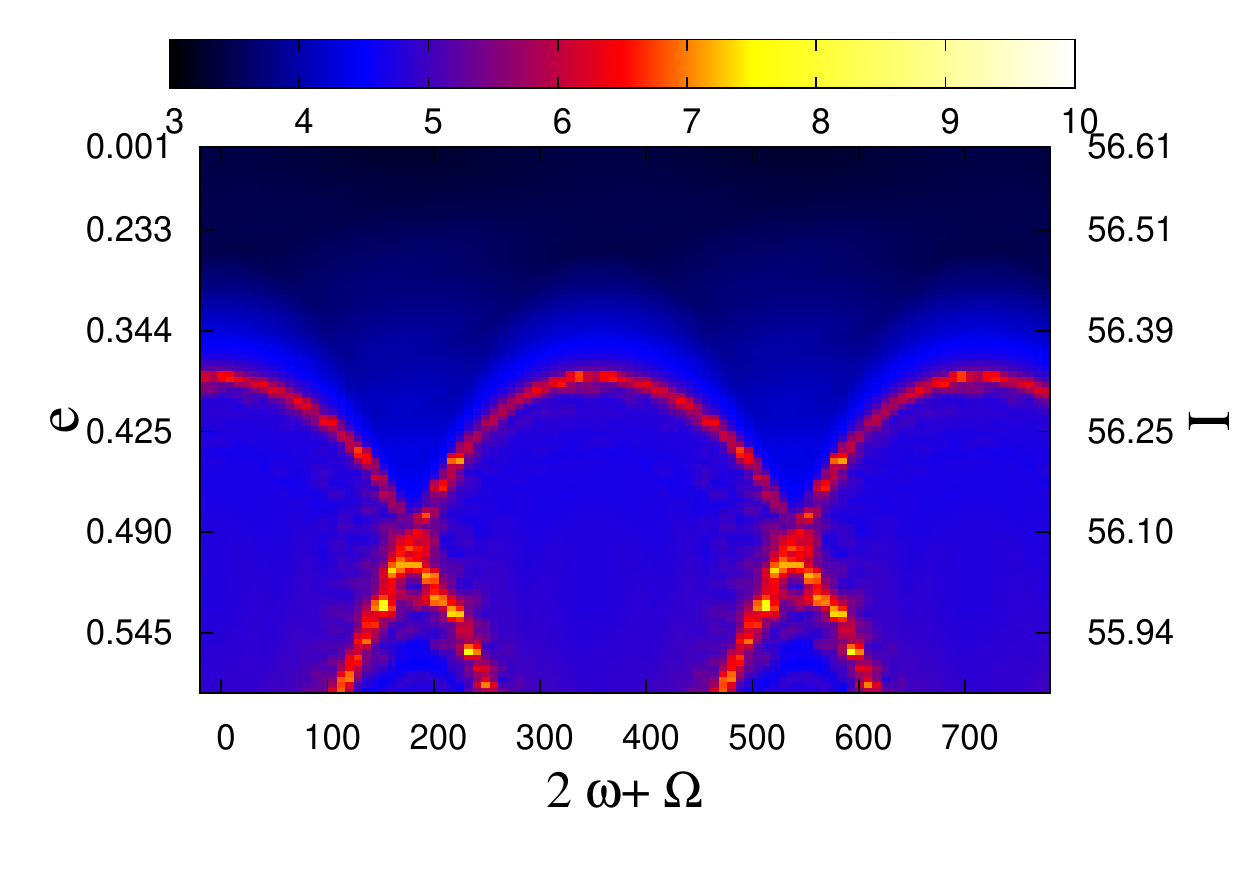}
\vglue0.5cm
\caption{FLIs for the model c), for $a=15\,000$ km, $\Omega=180^\circ$ and  $T=0.03$.}
\label{fig:FLI_model_c_a=15000}
\end{figure}

\begin{figure}[h]
\centering
\vglue0.1cm
\hglue0.1cm
\includegraphics[width=5truecm,height=4truecm]{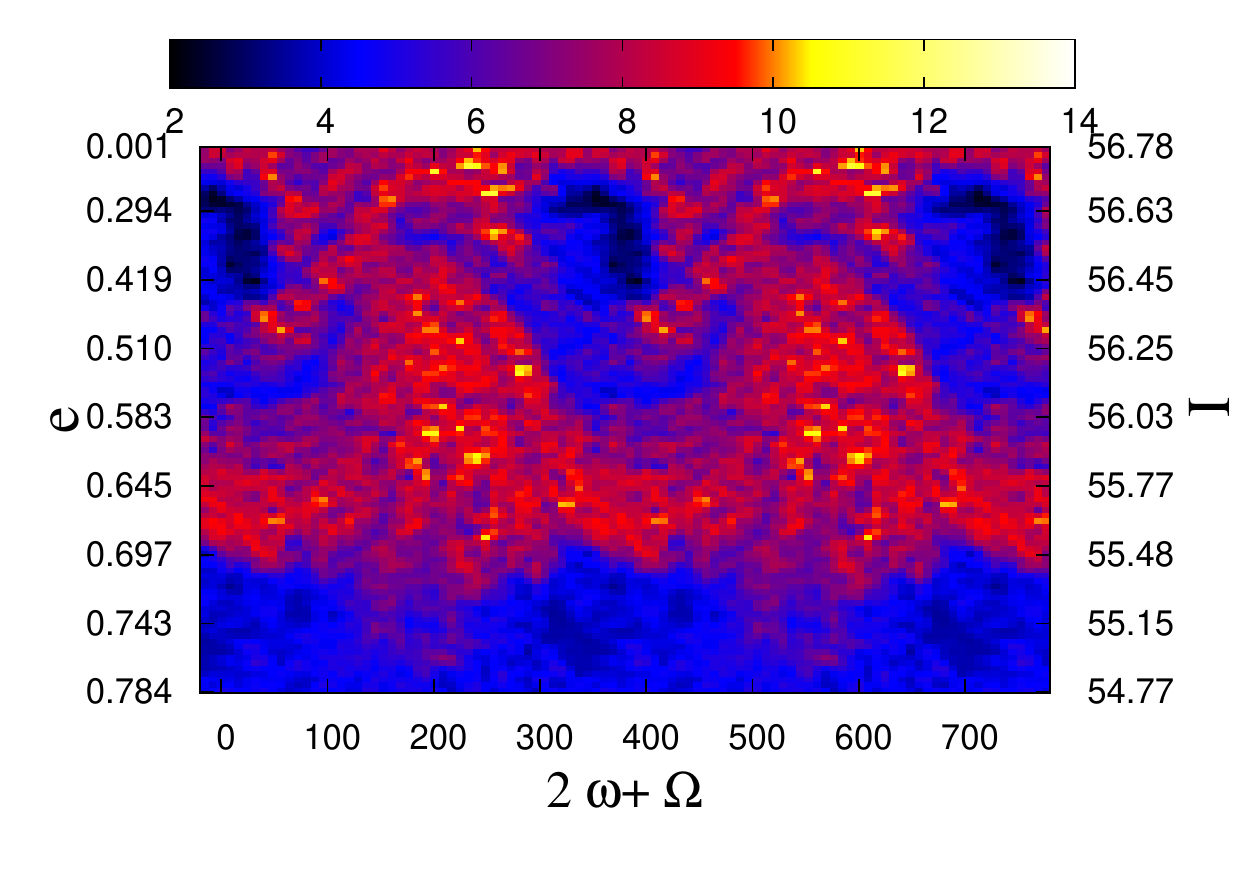}
\includegraphics[width=5truecm,height=4truecm]{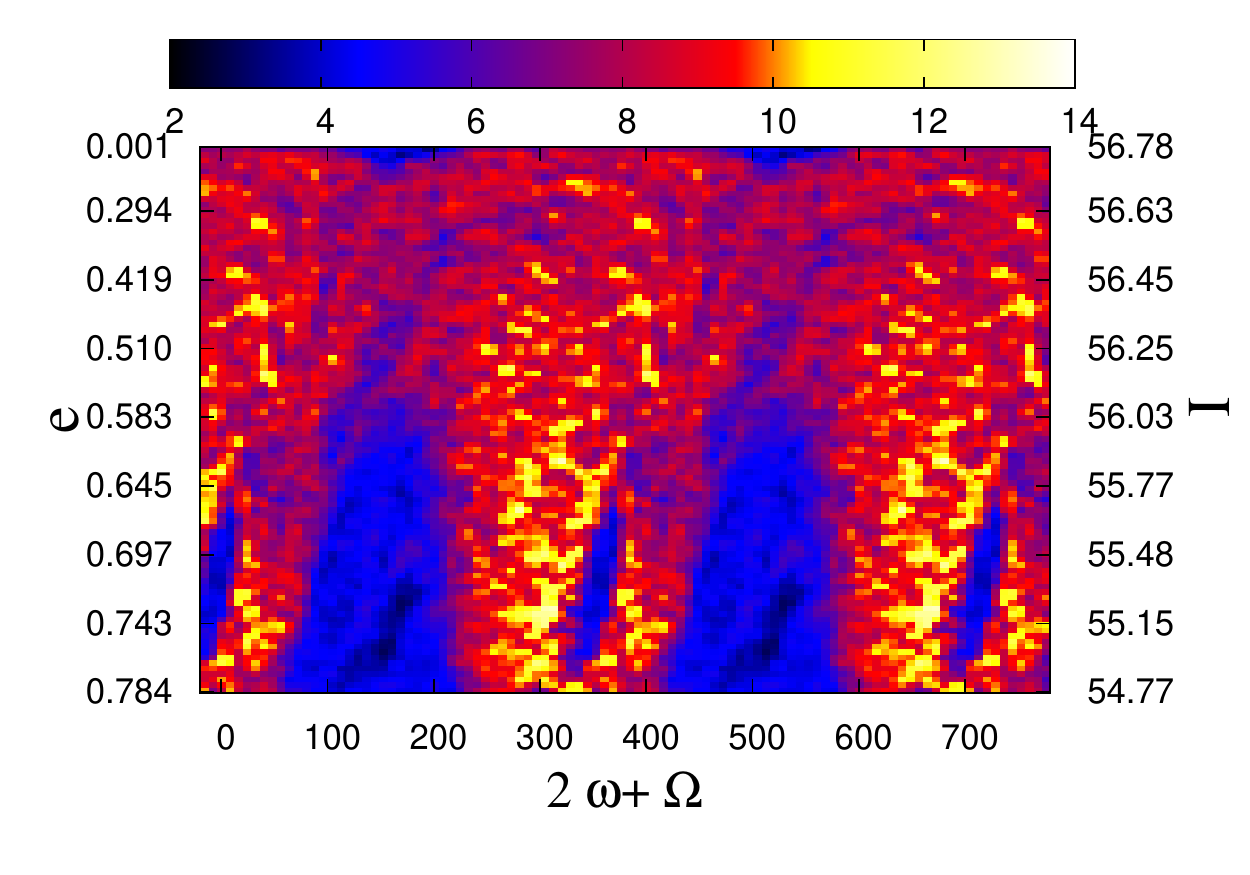}
\includegraphics[width=5truecm,height=4truecm]{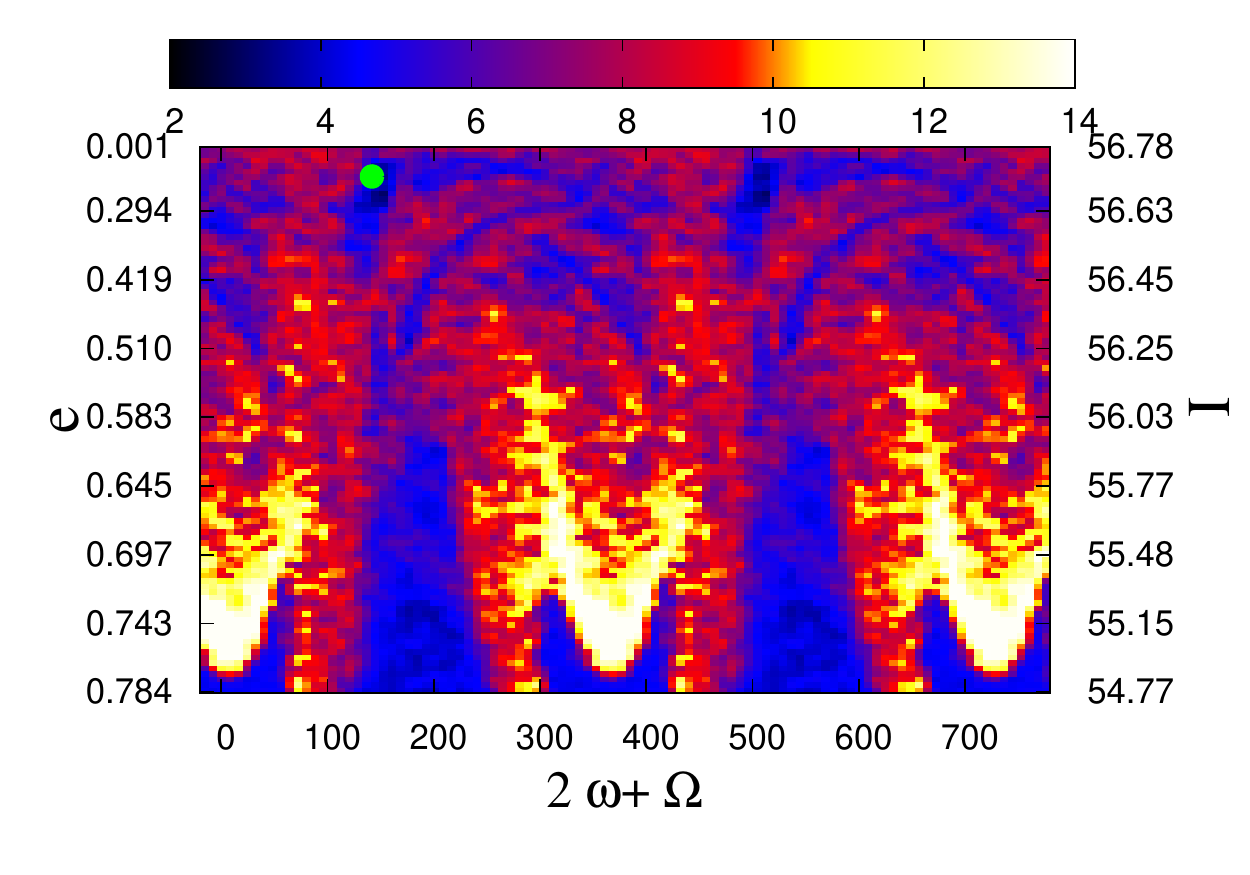}
\vglue0.5cm
\caption{FLIs for the model c), for $a=29\,546$ km $T=0.04$ and: $\Omega=0^\circ$ (left); $\Omega=90^\circ$ (middle); $\Omega=180^\circ$
(right). The green circle in the right plot represents an orbit analyzed in Figure~\ref{fig:orbits_T=0_04_Om=180_model_c}.}
\label{fig:FLI_model_c_Om=0_90_180}
\end{figure}

\begin{figure}[h]
\centering
\vglue0.1cm
\hglue0.1cm
\includegraphics[width=5truecm,height=4truecm]{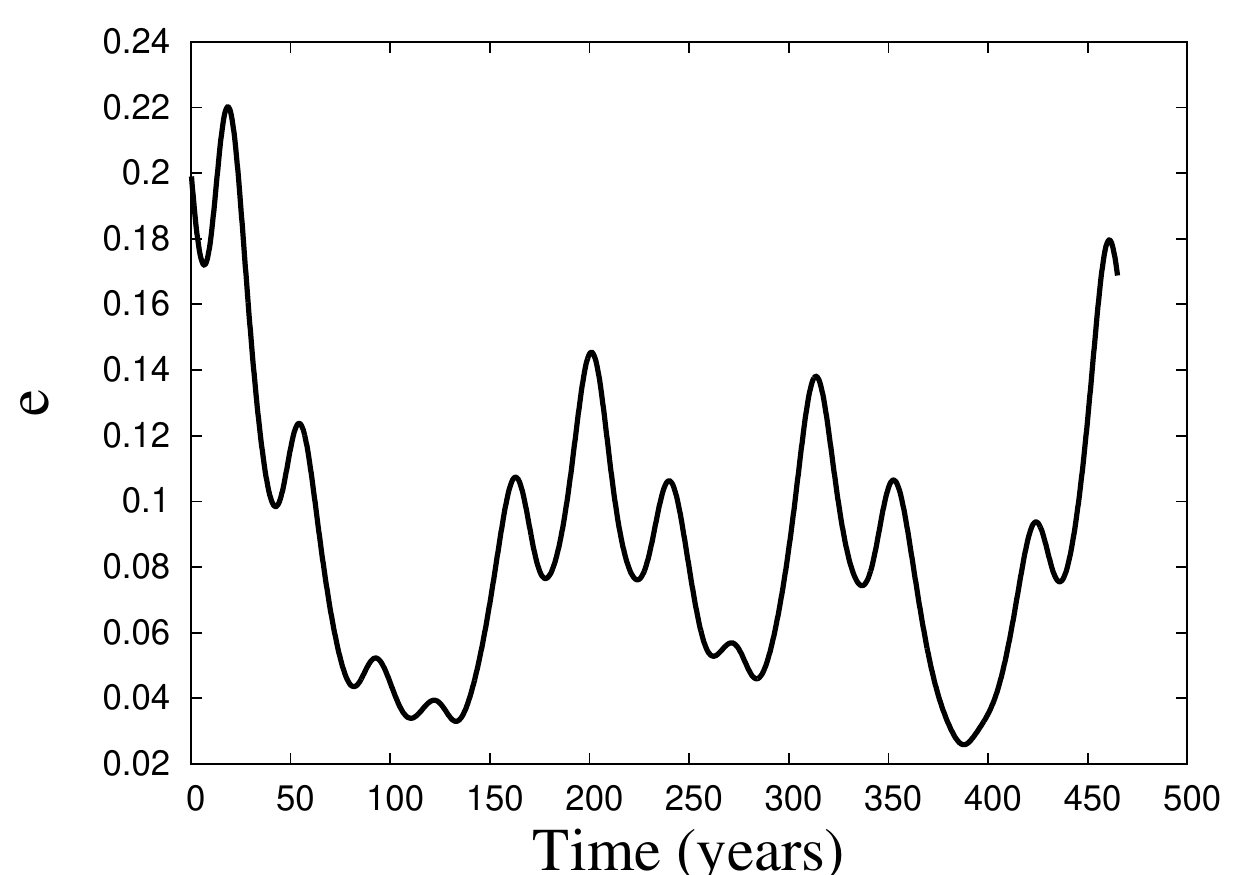}
\includegraphics[width=5truecm,height=4truecm]{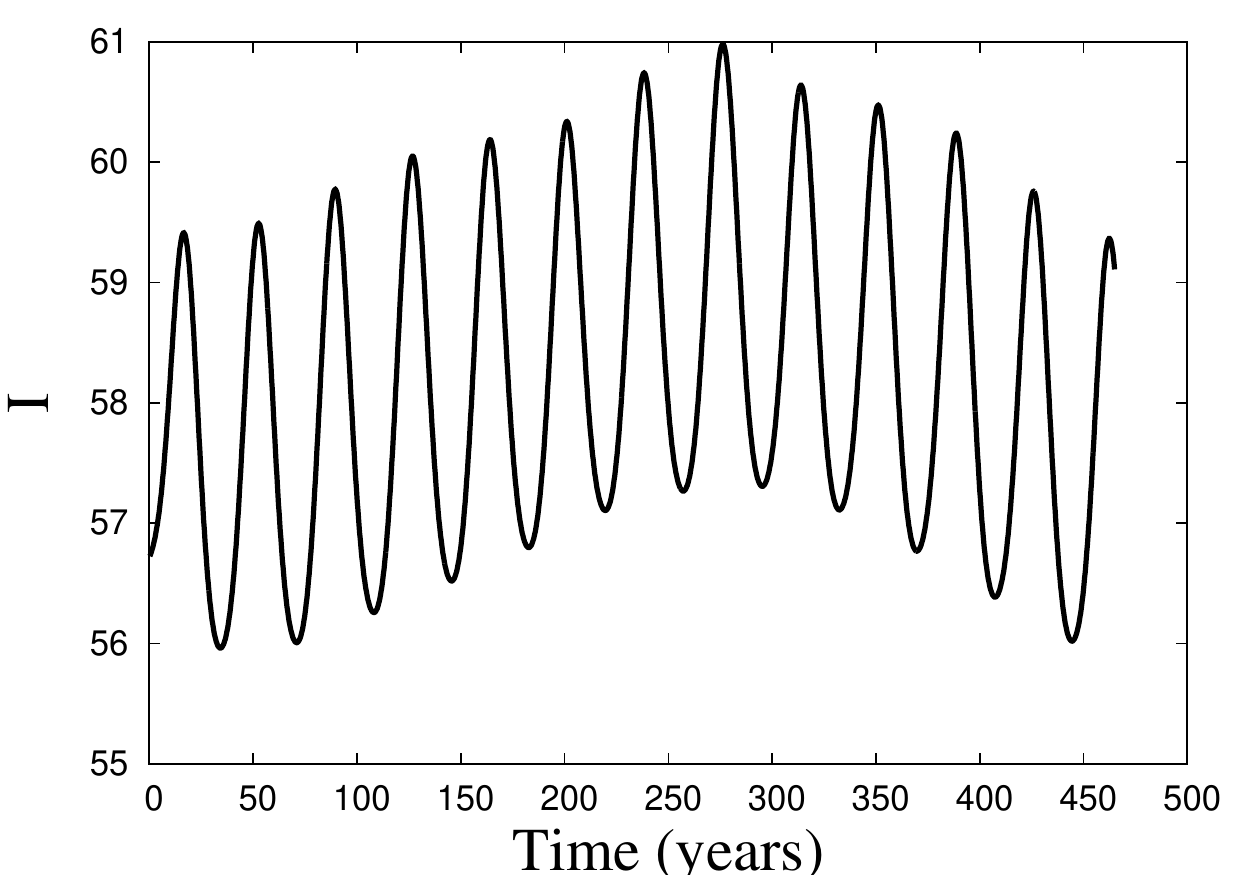}
\includegraphics[width=5truecm,height=4truecm]{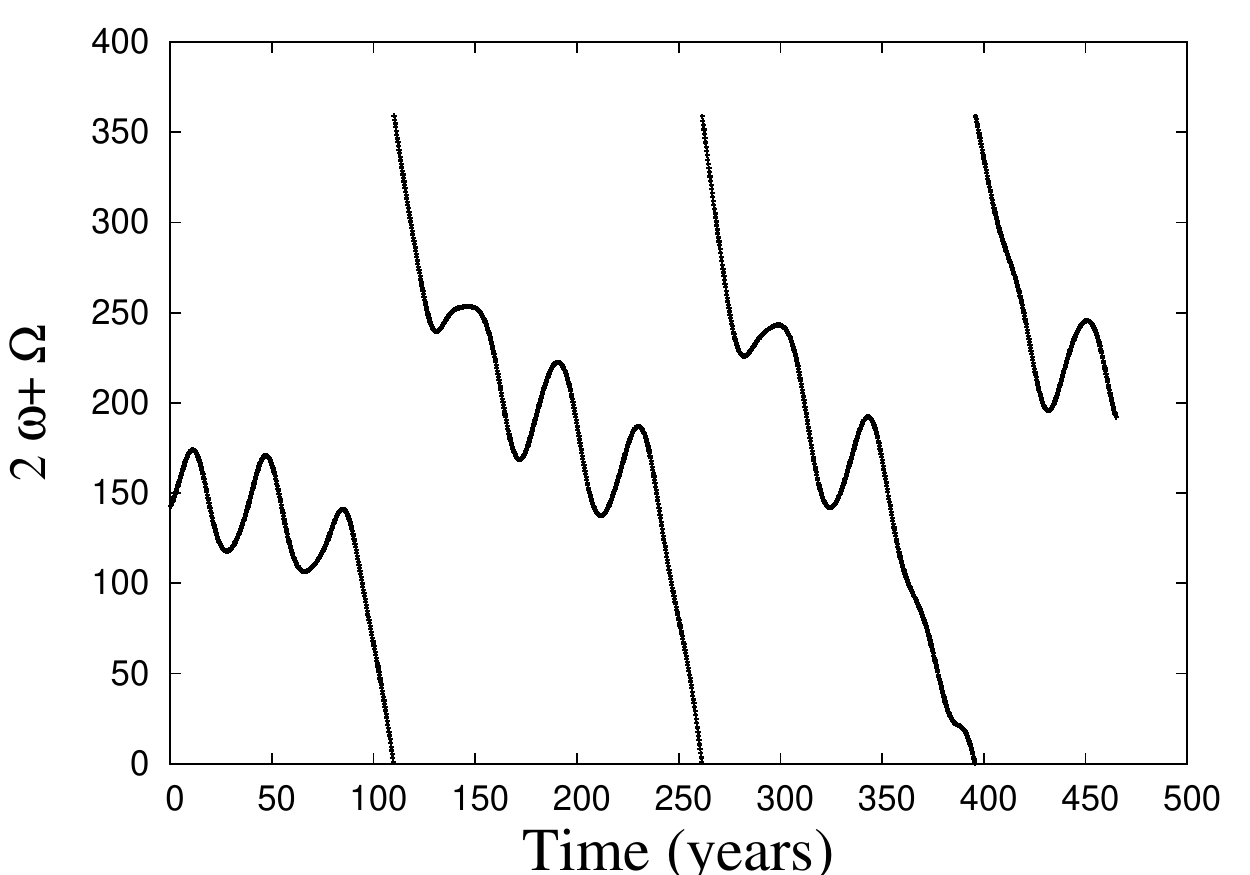}
\vglue0.5cm
\caption{Integration of the orbit having the initial conditions $T=0.04$, $\Omega=180^\circ$, $\sigma=142^\circ$ and $S=0.41$ (or $e=0.201$,
$I=56.71^\circ$).}
\label{fig:orbits_T=0_04_Om=180_model_c}
\end{figure}

We finally consider the dynamics associated to the more complete model c),
which is described by the non--autonomous Hamiltonian $\H$ introduced in \eqref{Hamiltonian}
The results are presented in Figures~\ref{fig:FLI_model_c_a=15000}--\ref{fig:orbits_T=0_04_Om=180_model_c}. As we already remarked above,
for $a=15\,000$ km, the phase plane $\sigma$--$S$ is very similar to the one described by model a), compare
Figure~\ref{fig:FLI_model_c_a=15000} with the left panel of Figure~\ref{fig:model_a}. However, for large $a$, the dynamics is much more
complex. Roughly speaking, on the global dynamical background described by model b), and which does not change significantly in a
vicinity of several km from the nominal distance of $a=29\,546$ km, one should superimpose the exact resonances shown in different colors in
the right bottom panel of Figure~\ref{WEB_structure}. These resonances are due to the variation of the lunar node, as noted by
\cite{ElyHowell}, \cite{aR15},  and their location depends on the value of the semimajor axis.

As a consequence, since the resonance $2\dot{\omega}+\dot{\Omega}=0$ is crossed by multiple exact resonances, having different widths (see
\cite{DRADVR15}), the orbital elements vary chaotically. One gets large regions filled by chaotic motions, marked by larger yellow-red values
of the FLI. In contrast with the model b), here the FLI values vary on a longer scale, from 2 to 14.
Figure~\ref{fig:FLI_model_c_Om=0_90_180} shows the results for $T=0.04$ and for $\Omega=0^\circ$ (left), $\Omega=90^\circ$ (middle) and
$\Omega=180^\circ$ (right). Comparing these plots with the corresponding ones obtained for model b), we remark that, besides the large
yellow-red regions obtained as effect of the overlapping of resonances (either the superposition of the exact resonances shown in the right
bottom panel of Figure~\ref{WEB_structure} with the exact resonance $2\dot{\omega}+\dot{\Omega}=0$, or with the critical inclination
resonance $\dot{\omega}=0$), some blue regions are noticeable, which account for the libration regions associated to the equilibrium points.
For instance, in the left plot of Figure~\ref{fig:FLI_model_c_Om=0_90_180}, we have a stable equilibrium point at about $\sigma=360^\circ$
and $e=0.294$ with a libration island (blue color) small in width (compare also with the left plot of
Figure~\ref{fig:FLI_model_b_Om=0_90_270}). Numerical tests show that an initial condition inside this region remains there, even if the
variations of $e$ and $\sigma$ are not regular.  The red-yellow regions visible for eccentricities larger than $0.5$ are due to the
interaction of the exact resonances depicted in Figure~\ref{WEB_structure}, bottom right plot, with the critical inclination resonance.

In both the middle and right panels of Figure~\ref{fig:FLI_model_c_Om=0_90_180}, we notice two important blue (libration) regions: one at
small eccentricities (the orbit marked with a green circle in the right panel of Figure~\ref{fig:FLI_model_c_Om=0_90_180} and analyzed in
Figure~\ref{fig:orbits_T=0_04_Om=180_model_c} is within this region) and one at large eccentricities (at about $\sigma=360^\circ$ and
$e=0.784$ in the right panel of Figure~\ref{fig:FLI_model_c_Om=0_90_180}). These regions show that the bifurcation phenomenon described by
the model a) is still valid for the more complete model c).

As a final remark, one should clarify what is happening inside the yellow-red region, for example in the middle panel of
Figure~\ref{fig:FLI_model_c_Om=0_90_180}. The answer is the following: usually one obtains an irregular growth in eccentricity. The
growth is due,  in essence, to the resonance $2 \dot{\omega}+ \dot{\Omega}=0$ (as the models a) and b) infer) and the irregular (chaotic)
behavior is obtained as an effect of the overlapping of the resonance $2\dot{\omega}+\dot{\Omega}=0$ with the resonances shown in
Figure~\ref{WEB_structure}.
We made several other experiments and found that colliding orbits can occur as a byproduct of the
eccentricity growth due to the interaction with the resonance $2\dot{\omega}+\dot{\Omega}=0$: the increase
of the eccentricity leads to have a distance at perigee less than the Earth's radius.
On the other hand, initial data in a chaotic region can undergo the effect of the interaction between different
resonance, but without leading to collisions.

\section{Conclusions}\label{sec:conclusions}
Lunisolar resonances might contribute to shape the dynamics of small bodies around the Earth
(\cite{sB01}, \cite{DRADVR15}, \cite{aR15}). Among such resonances, that corresponding to
$2\dot\omega+\dot\Omega=0$ is responsible for the growth in eccentricity. To explain this phenomenon,
we compare three different models with increasing complexity, obtained averaging over fast angles (model a)),
or just by neglecting the rate of variation of $\Omega_M$ (model b)), or rather including the variation of
$\Omega_M$ (model c)). A comparison among these models provide us with the ingredients which lead to chaos
and which provide an increase of the eccentricity.

By comparing the results of models a)-b)-c), we infer that the dynamics around the stable
equilibria at large values of the eccentricity is well represented by all models. On the
contrary, for small values of the eccentricity the effect of the variation of the lunar
longitude of the node plays a relevant role and, even if it occurs on long time scales, cannot
be neglected for an accurate description of the dynamics.

Finally, it is worth noticing that the growth in eccentricity provoked by the resonance $2\dot\omega+\dot\Omega=0$
can be used as an effective strategy to move space debris into non-operative or graveyard orbits.

\appendix

\section{Expressions for the lunar and solar Hamiltonians}\label{app:rmoonsun}

We report below the explicit expressions for $\H_{Moon}$ and $\H_{Sun}$.
\beqano
\H_{Moon}&=& -10^{-6} R_{Moon}\ ,\nonumber\\
\H_{Sun}&=& -10^{-6} R_{Sun}\ ,\nonumber\\
\eeqano
where
\beqano
R_{Moon}&=& 3.06238 a^2+4.59357 a^2 e^2+0.595633 a^2 e^2 \cos(2 \omega -2 \Omega )\nonumber\\
&-&
1.19127 a^2 e^2 \cos(I) \cos(2 \omega -2
\Omega )+0.595633 a^2 e^2 \cos^2(I) \cos(2 \omega -2 \Omega )\nonumber\\
&+&
0.476507 a^2 \cos(2 \Omega )+0.71476 a^2 e^2 \cos(2
\Omega )-0.476507 a^2 \cos^2(I) \cos(2 \Omega )\nonumber\\
&-&
0.71476 a^2 e^2 \cos^2(I) \cos(2 \Omega )+0.595633 a^2
e^2 \cos(2 \omega +2 \Omega )\nonumber\\
&+&
1.19127 a^2 e^2 \cos(I) \cos(2 \omega +2 \Omega )+0.595633 a^2 e^2 \cos^2(I)
\cos(2 \omega +2 \Omega )\nonumber\\
&+&
0.0000543 a^2 e^2 \cos(2 \omega -2 \Omega -2 \Omega_M)-0.0001086 a^2 e^2 \cos(I)
\cos(2 \omega -2 \Omega -2 \Omega_M)\nonumber\\
&+&
0.00005433 a^2 e^2 \cos^2(I) \cos(2 \omega -2 \Omega -2
\Omega_M)+0.02347 a^2 \cos(2 \Omega -2 \Omega_M)
\eeqano

\beqano
&+&
0.035207 a^2 e^2 \cos(2 \Omega -2 \Omega_M)-0.0234714
a^2 \cos^2(I) \cos(2 \Omega -2 \Omega_M)\nonumber\\
&-&
0.03520 a^2 e^2 \cos^2(I) \cos(2 \Omega -2
\Omega_M)+0.0293392 a^2 e^2 \cos(2 \omega +2 \Omega -2 \Omega_M)\nonumber\\
&+&
0.0586784 a^2 e^2 \cos(I) \cos(2
\omega +2 \Omega -2 \Omega _M)+0.0293392 a^2 e^2 \cos^2(I) \cos(2 \omega +2 \Omega -2 \Omega _M)\nonumber\\
&-&
0.011402
a^2 e^2 \cos(2 \omega -2 \Omega -\Omega_M)+0.0228039 a^2 e^2 \cos(I) \cos(2 \omega -2 \Omega -\Omega_M)\nonumber\\
&-&
0.011402 a^2 e^2 \cos^2(I) \cos(2 \omega -2 \Omega -\Omega_M)+0.211959 a^2 \cos(2 \Omega
- \Omega_M)\nonumber\\
&+&
0.317939 a^2 e^2 \cos(2 \Omega - \Omega_M)-0.211959 a^2 \cos^2(I) \cos(2 \Omega
- \Omega_M)\nonumber\\
&-&
0.317939 a^2 e^2 \cos^2(I)\cos(2 \Omega - \Omega_M)+0.264949 a^2 e^2 \cos(2
\omega +2 \Omega - \Omega_M)\nonumber\\
&+&
0.529898 a^2 e^2 \cos(I) \cos(2 \omega +2 \Omega - \Omega_M)+0.264949
a^2 e^2 \cos^2(I) \cos(2 \omega +2 \Omega - \Omega_M)\nonumber\\
&-&
0.405675 a^2 \cos(\Omega_M)-0.608513
a^2 e^2 \cos(\Omega_M)+0.00404032 a^2 \cos(2 \Omega_M)\nonumber\\
&+&0.00606 a^2 e^2 \cos(2 \Omega_M)
-0.00912157
a^2 \cos(2 \Omega +\Omega_M)\nonumber\\
&-&
0.01368 a^2 e^2 \cos(2 \Omega +\Omega_M)+0.009121 a^2 \cos^2(I)
\cos(2 \Omega +\Omega_M)\nonumber\\
&+&
0.0136823 a^2 e^2 \cos^2(I) \cos(2 \Omega +\Omega_M)-0.011402 a^2 e^2
\cos(2 \omega +2 \Omega +\Omega_M)\nonumber\\
&-&
0.0228039 a^2 e^2 \cos(I) \cos(2 \omega +2 \Omega +\Omega_M)-0.011402
a^2 e^2 \cos^2(I) \cos(2 \omega +2 \Omega +\Omega_M)\nonumber\\
&+&
0.264949 a^2 e^2 \cos(2 \omega -2\Omega +
\Omega_M)-0.529898 a^2 e^2 \cos(I) \cos(2 \omega -2 \Omega + \Omega_M)\nonumber\\
&+&
0.264949 a^2 e^2 \cos^2(I)
\cos(2 \omega -2 \Omega + \Omega_M)+0.0293392 a^2 e^2 \cos(2 \omega -2 \Omega +2 \Omega_M)\nonumber\\
&-& 0.0586784 a^2
e^2 \cos(I) \cos(2 \omega -2 \Omega +2 \Omega_M)+0.0293392 a^2 e^2 \cos^2(I) \cos(2
\omega -2 \Omega +2 \Omega_M)\nonumber\\
&+&
0.0000434 a^2 \cos(2 \Omega +2 \Omega_M)+0.0000652 a^2 e^2 \cos(2 \Omega
+2 \Omega_M)\nonumber\\
&-&
0.0000434 a^2 \cos^2(I) \cos(2 \Omega +2 \Omega_M)-0.0000652 a^2 e^2 \cos^2(I)
\cos(2 \Omega +2 \Omega_M)\nonumber\\
&+&
0.0000543 a^2 e^2 \cos(2 \omega +2 \Omega +2 \Omega_M)+0.0001086 a^2 e^2
\cos(I) \cos(2 \omega +2 \Omega +2 \Omega_M)\nonumber\\
&+&
0.0000543 a^2 e^2 \cos^2(I) \cos(2
\omega +2 \Omega +2 \Omega_M)+5.49537 a^2 e^2 \cos(2 \omega -\Omega ) \sin(I)\nonumber\\
&-&
5.49537 a^2 e^2 \cos(I)
\cos(2 \omega -\Omega ) \sin(I)+4.39629 a^2 \cos(I) \cos(\Omega ) \sin(I)\nonumber\\
&+&
6.59444 a^2 e^2
\cos(I) \cos(\Omega ) \sin(I)-5.49537 a^2 e^2 \cos(2 \omega +\Omega ) \sin(I)\nonumber\\
&-&
5.49537 a^2
e^2 \cos(I) \cos(2 \omega +\Omega ) \sin(I)+0.00104769 a^2 e^2 \cos(2 \omega -\Omega -2 \Omega_M) \sin(I)\nonumber\\
&-&
0.00104769 a^2 e^2 \cos(I) \cos(2 \omega -\Omega -2 \Omega_M) \sin(I)-0.0194763
a^2 \cos(I) \cos(\Omega -2 \Omega_M) \sin(I)\nonumber\\
&-&
0.0292145 a^2 e^2 \cos(I) \cos(\Omega
-2 \Omega_M) \sin(I)+0.0243454 a^2 e^2 \cos(2 \omega +\Omega -2 \Omega _M) \sin(I)\nonumber\\
&+&
0.0243454
a^2 e^2 \cos(I) \cos(2 \omega +\Omega -2 \Omega_M) \sin(I)-0.162524 a^2 e^2 \cos(2
\omega -\Omega -\Omega_M) \sin(I)\nonumber\\
&+&
0.162524 a^2 e^2 \cos(I) \cos(2 \omega -\Omega -\Omega_M)
\sin(I)+0.8898 a^2 \cos(I) \cos(\Omega - \Omega_M) \sin(I)\nonumber\\
&+&
1.33475 a^2 e^2 \cos(I)
\cos(\Omega - \Omega_M) \sin(I)-1.1123 a^2 e^2 \cos(2 \omega +\Omega - \Omega_M) \sin(I)\nonumber\\
&-&
1.1123
a^2 e^2 \cos(I) \cos(2 \omega +\Omega - \Omega_M) \sin(I)-0.130019 a^2 \cos(I)
\cos(\Omega +\Omega_M) \sin(I)\nonumber\\
&-&
0.19502 a^2 e^2 \cos(I) \cos(\Omega +\Omega_M)
\sin(I)+0.16252 a^2 e^2 \cos(2 \omega +\Omega +\Omega_M) \sin(I)\nonumber\\
&+&
0.162524 a^2 e^2 \cos(I)
\cos(2 \omega +\Omega +\Omega_M) \sin(I)+1.1123 a^2 e^2 \cos(2 \omega -\Omega + \Omega_M)
\sin(I)\nonumber\\
&-&
1.1123 a^2 e^2 \cos(I) \cos(2 \omega -\Omega + \Omega_M) \sin(I)-0.02434
a^2 e^2 \cos(2 \omega -\Omega +2 \Omega_M) \sin(I)\nonumber\\
&+&
0.0243454 a^2 e^2 \cos(I) \cos(2
\omega -\Omega +2 \Omega_M) \sin(I)+0.000838 a^2 \cos(I) \cos(\Omega +2 \Omega_M) \sin(I)\nonumber\\
&+&
0.00125723
a^2 e^2 \cos(I) \cos(\Omega +2 \Omega_M) \sin(I)
-0.00104769 a^2 e^2 \cos(2 \omega +\Omega
+2 \Omega_M) \sin(I)\nonumber\\
&-&
0.00104769 a^2 e^2 \cos(I) \cos(2 \omega +\Omega +2 \Omega_M) \sin(I)
-4.59357
a^2 \sin^2(I)
\eeqano

\beqano
&-&
6.89035 a^2 e^2 \sin(I)^2+11.4839 a^2 e^2 \cos(2 \omega ) \sin^2(I)+0.00757559
a^2 e^2 \cos(2 \omega -2 \Omega_M) \sin^2(I)\nonumber\\
&-&
0.760641 a^2 e^2 \cos(2 \omega -\Omega_M) \sin^2(I)+0.608513
a^2 \cos(\Omega_M) \sin^2(I)\nonumber\\
&+&
0.912769 a^2 e^2 \cos(\Omega_M) \sin^2(I)-0.00606
a^2 \cos(2 \Omega_M) \sin^2(I)-0.00909 a^2 e^2 \cos(2 \Omega_M) \sin^2(I)\nonumber\\
&-&
0.760641
a^2 e^2 \cos(2 \omega +\Omega_M) \sin^2(I)+0.007575 a^2 e^2 \cos(2 \omega +2 \Omega_M) \sin^2(I)\ ,
\eeqano

\beqano
R_{Sun}&=&1.42243 a^2 + 2.13364 a^2 e^2 + 0.22133 a^2 \cos(2\Omega) +
 0.331995 a^2 e^2 \cos(2\Omega) \nonumber\\
 &-&
 0.22133 a^2 \cos(2\Omega) \cos^2(I) -
 0.331995 a^2 e^2 \cos(2\Omega) \cos^2(I) +
 0.276662 a^2 e^2 \cos(2\Omega - 2 \omega) \nonumber\\
 &-&
 0.553324 a^2 e^2 \cos(I) \cos(2\Omega - 2 \omega) +
 0.276662 a^2 e^2 \cos^2(I) \cos(2\Omega - 2\omega) \nonumber\\
 &+&
 0.276662 a^2 e^2 \cos(2\Omega + 2\omega) +
 0.553324 a^2 e^2 \cos(I) \cos(2\Omega + 2\omega) \nonumber\\
 &+&
 0.276662 a^2 e^2 \cos^2(I) \cos(2\Omega + 2\omega) +
 2.04201 a^2 \cos(\Omega) \cos(I) \sin(I) \nonumber\\
 &+&
 3.06301 a^2 e^2 \cos(\Omega) \cos(I) \sin(I) +
 2.55251 a^2 e^2 \cos(\Omega - 2\omega) \sin(I) \nonumber\\
 &-&
 2.55251 a^2 e^2 \cos(I) \cos(\Omega - 2\omega) \sin(I) -
 2.55251 a^2 e^2 \cos(\Omega + 2\omega) \sin(I) \nonumber\\
 &-&
 2.55251 a^2 e^2 \cos(I) \cos(\Omega + 2\omega) \sin(I) -
 2.13364 a^2 \sin(I)^2 - 3.20046 a^2 e^2 \sin^2(I) \nonumber\\
 &+&
 5.33411 a^2 e^2 \cos(2\omega) \sin^2(I)\ .
\eeqano

\section{Fast Lyapunov Indicators}\label{app:FLI}
The FLIs were introduced in \cite{froes} as the largest Lyapunov characteristic exponents
at a given time, say $t=T$. Their definition is the following. Consider the $n$--dimensional differential system
$$
\dot{{\bf x}}={\bf F}({\bf x})
$$
with ${\bf x}\in\real^n$. Let the corresponding variational equations be
$$
\dot{{\bf v}}=\Big({{\partial {\bf F}({\bf x})} \over {\partial {\bf x}}}\Big)\ {\bf v}\ ,
$$
where ${\bf v}\in\real^n$. Consider the initial conditions ${\bf x}(0) \in \real^n$,
${\bf v}(0) \in \real^n$; the FLI at time $T\geq 0$ is defined as
$$
{\rm FLI}({\bf x}(0), {\bf v}(0), T) \equiv \sup _{0 < t\leq T} \log ||{\bf v}(t)||\  .
$$
Small values of FLIs correspond to regular (periodic or quasi--periodic) dynamics,
while large values denote chaotic motions.

\bibliographystyle{spmpsci}

\begin{thebibliography}{}



\bibitem[Breiter(2001)]{sB01}
Breiter, S.: Lunisolar resonances revisited. Celest. Mech. Dyn. Astr. {\bf 81}, 81--91 (2001)

\bibitem[Celletti and Gale\c{s}(2014)]{CGmajor}
Celletti, A., Gale\c{s}, C.:
On the dynamics of space debris: 1:1 and 2:1 resonances.
J. Nonlinear Science \textbf{24}, 1231--1262 (2014)

\bibitem[Celletti and Gale\c{s}(2015)]{CGext}
Celletti, A. Gale\c{s}, C.:
A study of the main resonances outside the geostationary ring.
Adv. Space Res. \textbf{56}, 388--405 (2015)

\bibitem[Celletti et al.(2016a)] {CGPbif}
Celletti, A.,  Gale\c s, C., Pucacco:  Bifurcation of lunisolar secular resonances for space debris orbits.
arXiv:1512.02178 (2016a)

\bibitem[Celletti et al.(2016b)] {CGPRnote}
Celletti, A.,  Gale\c s, C., Pucacco, G.,   Rosengren, A.:  Analytical development of the lunisolar disturbing function and the critical inclination secular resonance. (submitted) (2016b)

\bibitem[Chirikov(1979)]{chirikov}
Chirikov, B.V.: A universal instability of many-dimensional oscillator systems.
Phys. Rep. \textbf{52}, 263--379 (1979)

\bibitem[Cook(1962)]{gC62}
Cook, G.E.:
Luni-solar perturbations of the orbit of an Earth satellite.
Geophys. J. \textbf{6}, 271--291 (1962)

\bibitem[Daquin et al.(2016)]{DRADVR15}
Daquin J., Rosengren A. J., Alessi E. M., Deleflie F., Valsecchi G. B., Rossi A., 2015: The dynamical structure of the MEO region: long-term
stability, chaos, and transport. Celest. Mech. Dyn. Astr., doi:10.1007/s10569-015-9665-9 (2016)

\bibitem[Ely and Howell(1997)]{ElyHowell}
Ely, T.A.,   Howell, K.C.:  Dynamics of artificial satellite orbits with tesseral resonances including the effects of luni--solar
perturbations. Dynamics and Stability of Systems {\bf 12},  243--269 (1997)

\bibitem[Froeschl\'e et al.(1997)]{froes}
Froeschl\'e, C.,   Lega, E.,   Gonczi, R.:  Fast Lyapunov indicators. Application to asteroidal motion. Celest. Mech. Dyn. Astr. {\bf 67}, 41--62 (1997)

\bibitem[Hughes(1980)]{HughesI}
Hughes, S.:
Earth satellite orbits with resonant lunisolar perturbations. I. Resonances dependent only on inclination.
Proc. R. Soc. Lond. A \textbf{372}, 243--264 (1980)

\bibitem[Kaula(1962)]{wK62}
Kaula, W.M.:
Development of the lunar and solar disturbing functions for a close satellite.
Astron. J. \textbf{67}, 300--303 (1962)

\bibitem[Lane(1989)]{mL89}
Lane, M.T.:
On analytic modeling of lunar perturbations of artificial satellites of the Earth.
Celest. Mech. Dyn. Astr. \textbf{46}, 287--305 (1989)

\bibitem[Radtke et al.(2015)]{RDGF2015}
Radtke, J., Dominguez-Gonzalez, R.,  Flegel, S.K.,  Sanchez-Ortiz, N.,  Merz, K.: Impact of eccentricity build-up and graveyard disposal Strategies on MEO navigation constellations.  Adv. Space Res.
\textbf{56}, 2626--2644 (2015)

\bibitem[Rosengren et al.(2015a)]{aR15}
Rosengren, A.J., Alessi, E.M., Rossi, A., Valsecchi, G.B.:
Chaos in navigation satellite orbits caused by the perturbed motion of the Moon.
Mon. Not. R. Astron. Soc. doi:10.1093/mnras/stv534 (2015a)

\bibitem[Rosengren et al.(2015b)]{RDADRV15}
Rosengren, A.J., Daquin, J., Alessi, E.M., Deleflie, F., Rossi, A., Valsecchi, G.B.:
Galileo disposal strategy: stability, chaos and predictability. arXiv:1512:05822v1 (2015b)

\bibitem[Rossi(2008)]{aR08}
Rossi, A.:  Resonant dynamics of Medium Earth Orbits: space debris issues.
Celest. Mech. Dyn. Astr. {\bf 100}, 267--286 (2008)

\bibitem[Sanchez et al.(2015)]{Sanchez15}
Sanchez, D.M.,  Yokoyama, T.,  de Almeida Prado, A.F.B.: Study of some strategies for disposal of the GNSS satellites, Mathematical Problems
in Engineering. Volume {\bf 2015}, Article ID 382340, 14 pages (2015)


\end{thebibliography}

\end{document}